%% file: clas_hyperon_cxcz.tex
\documentclass[prc,superscriptaddress,showpacs,twocolumn,amssymb,amsmath,amsfonts,aps]{revtex4}

\usepackage{graphicx} 
\usepackage{dcolumn}  
\usepackage{longtable}
\usepackage{bm}       

\begin{document}


\input{authors}
\title{First Measurement of Beam-Recoil Observables 
 $C_x$ and $C_z$ in Hyperon Photoproduction.}


\date{\today}  


\begin{abstract} 

Spin transfer from circularly polarized real photons to recoiling
hyperons has been measured for the reactions $\vec\gamma + p \to K^+ +
\vec\Lambda$ and $\vec\gamma + p \to K^+ + \vec\Sigma^0$.  The data
were obtained using the CLAS detector at Jefferson Lab for
center-of-mass energies $W$ between 1.6 and 2.53 GeV, and for
$-0.85<\cos\theta_{K^+}^{c.m.}< +0.95$.  For the $\Lambda$, the
polarization transfer coefficient along the photon momentum axis,
$C_z$, was found to be near unity for a wide range of energy and kaon
production angles. The associated transverse polarization coefficient,
$C_x$, is smaller than $C_z$ by a roughly constant difference of unity.
Most significantly, the {\it total} $\Lambda$ polarization vector,
including the induced polarization $P$, has magnitude consistent with
unity at all measured energies and production angles when the beam is
fully polarized.  For the $\Sigma^0$ this simple phenomenology does
not hold.  All existing hadrodynamic models are in poor agreement with
these results.

\end{abstract}

\maketitle

\section{\label{intro}INTRODUCTION}

Photoproduction of strangeness off the proton leading to
$K\Lambda$ and $K\Sigma$ states is a fundamental process that is part
of the broader field of elementary pseudoscalar meson production.
It has been used primarily as a tool to investigate the formation and
decay of non-strange baryon resonances in a manner complimentary to
$\pi$ and $\eta$ meson production.  Spin observables such as those
reported here are expected to be sensitive tests of baryon resonance
structure and reaction models.  

When the photon beam is unpolarized, parity conservation in
electromagnetic production allows induced polarization, $P$, of the
hyperon only along the axis perpendicular to the reaction plane $\hat
\gamma \times \hat K$.  However, when the incoming photons are
circularly polarized, that is, when the photons are spin polarized
parallel or anti-parallel to the beam direction, giving them net
helicity, then this polarization may be transferred in whole or in
part to the spin orientation of the produced hyperons within the
reaction plane.  $C_x$ and $C_z$ characterize the polarization
transfer from a circularly polarized incident photon beam to a
recoiling hyperon along orthogonal axes in the reaction plane.  This
paper reports first measurements of the two double polarization
observables, $C_x$ and $C_z$, for $K^+ \Lambda$ and $K^+ \Sigma ^0$
photoproduction.  

Recent measurements of the photoproduction differential cross sections
have been published by groups working at Jefferson
Lab~\cite{bradforddsdo}, Bonn~\cite{bonn2}, and SPring-8~\cite{leps}.
Induced hyperon recoil polarizations, $P$, have also been published by
Jefferson Lab~\cite{mcnabb}, Bonn~\cite{bonn2}, and
GRAAL~\cite{graal}.  The beam linear polarization asymmetry, $\Sigma$,
was measured at SPring-8~\cite{zegers}.  These results were obtained
with large-acceptance detectors that allowed statistically precise
measurements across a broad range of kinematics.  Very sparse data
exist on the target asymmetry, $T$, from Bonn~\cite{althoff}.  A
preliminary version of the results reported in this paper was previously
given at the NStar 2005 conference \cite{nstar}.

Much of the recent experimental effort has been motivated by
theoretical calculations which suggest that strangeness
photoproduction might be a fertile place to search for non-strange
baryon resonances that couple strongly to $K^+ Y$ \cite{cap}.  Quark
model states ``missing'' in the analysis of single pion final states
of electromagnetic and hadronic production may merely be ``hidden''
due to unfavorable coupling strengths or complex multi-pion final
states. The less well studied strangeness production channels (as well
as other mesonic final states) cast a different light on the baryon
resonance spectrum.

The recently published differential cross sections have been tests for
a number of single channel theoretical models \cite{maid, mart, jan,
jan_a, saghai1, ireland, corthals}.  These models were mostly
tree-level calculations that attempted to extract information about
states decaying to $K^+ \Lambda$ or $K^+ \Sigma ^0$ by varying the
prescription for the inclusion of baryon resonances, the methods of
enforcing gauge invariance, and the introduction of hadronic form
factors, etc.  As the models were adjusted to the new differential
cross section measurements, there was a claim for evidence of a
specific new baryonic state \cite{maid} visible via $K^+\Lambda$
production.  However, it is clear that there is no unique solution for
the baryon resonance content of the differential and single
polarization observable data that is currently available
\cite{penner,jan,saghai1}.  Since the single channel models failed to
produce conclusive results for the baryon resonance content of
hyperon photoproduction, let alone undiscovered states, measurements of
new observables are needed in order to achieve better understanding
from $K^+ \Lambda$ and $K^+ \Sigma^0$.

Some more recent models have become more sophisticated by moving
beyond single channel analyses.  These fall into categories of either
coupled channel approaches~\cite{chiang,shklyar,saghai2} or of fitting
to multiple but independent reaction channels at once
~\cite{sarantsev,anisovich,anisovich2}.  On the side of greater
simplicity, one can compare the present results with a pure Regge
model~\cite{lag1, lag2} that contains no baryon resonance
contributions at all.  These models will be discussed and compared to
the present results later in this paper, however none of the models
will have been adjusted to fit the results presented here.

This paper will describe what $C_x$ and $C_z$ are and
how they are measured in Section \ref{sec:method}.  The experimental
setup will be outlined in Section \ref{sec:setup}, and specifics of
the data analysis will be covered in Section \ref{sec:analysis}.  The
results of the present measurements and discussion of what was found
will be given in Section \ref{sec:results}, including comparison to
predictions of seven different models.  Our conclusions will be
restated in Section \ref{sec:conclusions}.

\section{\label{sec:method}FORMALISM AND MEASUREMENT METHOD}

Real photoproduction of pseudoscalar mesons is fully described by four
complex amplitudes.  The bilinear combinations of these amplitudes
define 16 observables \cite{barker, knoechlein}, summarized in
Table~\ref{obstable}.  Of these sixteen observables, besides the
unpolarized differential cross section, there are three single
polarization observables and twelve double polarization observables.
The single polarization observables include the hyperon recoil
polarization ($P$), and the beam $(\Sigma)$ and target $(T)$
polarization asymmetries.  The double polarization observables
characterize reactions under various combinations of beam, target, and
baryon recoil polarization.  To uniquely determine the underlying
complex amplitudes, one has to measure the unpolarized cross section,
the three single polarization observables, and at least four double
polarization observables \cite{barker, tabakin1}.  To date, only $P$
and $\Sigma$ have been measured extensively and analyzed in models of
$K^+ \Lambda$ and $K^+ \Sigma ^0$ photoproduction.

\begin{table}
\begin{center}
\begin{tabular}{|c|c|c|c|}
\hline
& \multicolumn{3}{c|}{Required Polarization}\\
\raisebox{1.5ex}[0pt]{Observable} & Beam & Target & Hyperon\\
\hline
\hline
\multicolumn{4}{c}{Single Polarization \& Cross Section}\\
\hline
A, $\frac{d \sigma}{d \Omega}$ & - & - & -\\
$\Sigma$ & linear & - & -\\
$T$ & - & transverse & -\\
$P$ & - & - & along $y^\prime$\\
\hline
\hline
\multicolumn{4}{c}{Beam and Target Polarization}\\
\hline
$G$ & linear & along $z$ & -\\
$H$ & linear & along $x$ & -\\
$E$ & circular & along $z$ & -\\
$F$ & circular & along $x$ & -\\
\hline
\hline
\multicolumn{4}{c}{Beam and Recoil Baryon Polarization}\\
\hline
$O_{x'}$ & linear & - & along $x^\prime$\\
$O_{z'}$ & linear & - & along $z^\prime$\\
$C_{x'}$ & circular & - & along $x^\prime$\\
$C_{z'}$ & circular & - & along $z^\prime$\\
\hline
\hline
\multicolumn{4}{c}{Target and Recoil Baryon Polarization}\\
\hline
$T_{x'}$ & - & along $x$ & along $x^\prime$\\
$T_{z'}$ & - & along $x$ & along $z^\prime$\\
$L_{x'}$ & - & along $z$ & along $x^\prime$\\
$L_{z'}$ & - & along $z$ & along $z^\prime$\\
\hline
\end{tabular}
\caption{Groupings of all observables for pseudoscalar
  meson photoproduction.  The axis convention used in this paper to
  define alternate to the ``primed'' variables $C_{x^\prime}$ and
  $C_{z^\prime}$ are discussed in the text.  The table is adapted from
  Ref.~\cite{barker}.}
\label{obstable}
\end{center}
\end{table}

The present measurements were made with a circularly polarized photon
beam. Let $P_\odot$ represent the degree of beam polarization between
$-1.0$ and $+1.0$.  The spin-dependent cross section for $K^+Y$
photoproduction can be expressed as
\begin{equation}
\rho_Y \frac{d\sigma}{d\Omega_{K^+}} =
\left. {\frac{d\sigma}{d\Omega_{K^+}}}\right| _{unpol.} \!\!\!\!\!\!\!\!\!\!\!
\left\{ 1 + \sigma_y P + 
P_\odot (C_x \sigma_x + C_z \sigma_z ) \right \}.
\label{eq:cross}
\end{equation}
Here $\rho_Y$ is twice the density matrix of the ensemble of recoiling
hyperons $Y$ and is written
\begin{equation}
\rho_Y = (1 + \vec\sigma\cdot \vec P_Y),
\label{eq:density}
\end{equation}
where $\vec \sigma$ are the Pauli spin matrices and $\vec P_Y$ is the
measured polarization of the recoiling hyperons.  In Eq.~\ref{eq:cross}
the spin observables are the induced polarization $P$, and the
polarization transfer coefficients $C_x$ and $C_z$.  For further
discussion a definite coordinate system is needed.

\begin{figure}
\centering
\includegraphics[scale=0.35,angle=-90.0]{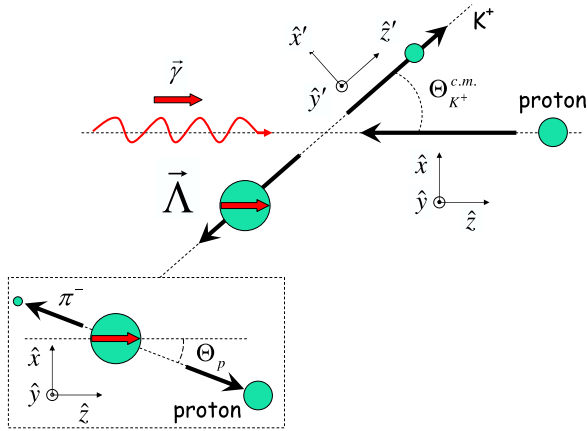}
\caption{ (Color online) 
In the overall reaction center of mass, the
coordinate system can be oriented along the outgoing $K^+$ meson
$\{\hat{x}^\prime,\hat{y}^\prime,\hat{z}^\prime\}$ or along the
incident photon direction $\{\hat{x},\hat{y},\hat{z}\}$.  The dotted
box represents the rest frame of the hyperon, and the coordinate
system used for specifying the polarization components. The red arrows
represent polarization vectors.}
\label{fig:axes}       
\end{figure}

Figure~\ref{fig:axes} illustrates the coordinate system used in this
paper.  In the literature there are two conventions for discussing the
beam-recoil observables.  The polarization of the hyperons in the
production plane can be described with respect to a $z$ axis chosen
along the incident beam direction ({\it i.e.} the helicity axis of the
photons) or along the momentum axis of the produced $K^+$.  Because
a polarization vector transforms as a vector in three-space, this
choice is of no fundamental significance.  In this paper we select
the $z$ axis along the photon helicity direction because it will be
seen that the transferred hyperon polarization is dominantly along
$\hat z$ defined in Fig.~\ref{fig:axes}.  Model calculations for $C_x$
and $C_z$ supplied to us in the $\{\hat x^\prime, \hat z^\prime\}$ basis
were rotated about the $\hat y$-axis to the $\{\hat x, \hat z\}$ basis.

With the axis convention chosen to give the results their simplest
interpretation, we correspondingly define our $C_x$ and $C_z$ with
signs opposite to the version of Eq.~\ref{eq:cross} given in
Ref.~\cite{barker}.  This will make $C_z$ positive when the $\hat z$
and $\hat z^\prime$ axes coincide at the forward meson production
angle, meaning that positive photon helicity results in positive
hyperon polarization along $\hat z$.

The connection between the measured hyperon recoil polarization $\vec
P_Y$ and the spin correlation observables $P$, $C_x$, and $C_z$, is
obtained by taking the expectation value of the spin operator $\vec\sigma$
with the density matrix $\rho_Y$ via the trace: $\vec P_Y = Tr(\rho_Y
\vec \sigma)$. This leads to the identifications
\begin{eqnarray}
P_{Y x} &=& P_\odot C_x 
\label{eq:obsx} \\
P_{Y y} &=& P 
\label{eq:obsp} \\
P_{Y z} &=& P_\odot C_z. 
\label{eq:obsz}
\end{eqnarray}
Thus, the transverse or induced polarization of the hyperon, $P_{Y y}$,
is equivalent to the observable $P$, while the $\hat x$ and $\hat z$
components of the hyperon polarization in the reaction plane are
proportional to $C_x$ and $C_z$ via the beam polarization factor
$P_\odot$.  Physically, $C_x$ and $C_z$ measure the transfer of
circular polarization, or helicity, of the incident photon on an
unpolarized target to the produced hyperon.

\subsection{\label{sec:asym} Hyperon Decay and Beam Helicity Asymmetries}

Hyperon polarizations $\vec P_Y$ are measured through the decay
angular distributions of the hyperons' decay products.  The decay $\Lambda \to
\pi^- p$ has a parity-violating weak decay angular distribution in the
$\Lambda$ rest frame.  The decay of the $\Sigma^0$ always proceeds
first via an $M1$ radiative decay to a $\Lambda$.  In either case,
$\vec{P}_Y$ is measured using the angular distribution of the decay
protons in the hyperon rest frame.  In the specified coordinate system
$i\in\{x,y,z\}$ is one of the three axes.  The decay distribution,
$I_i(\cos \theta_i)$, is given by
\begin{equation}
I_i(\cos \theta_i) = {\textstyle\frac{1}{2}}(1 + \nu \alpha P_{Y i} \cos \theta_i),
\label{eq:decay}
\end{equation}
where $\theta_i$ is the proton polar angle with respect to the given
axis in the hyperon rest frame.  The weak decay asymmetry, $\alpha$,
is taken to be $0.642$.  The factor $\nu$ is a ``dilution'' arising in
the $\Sigma^0$ case due to its radiative decay to a $\Lambda$, and
which is equal to $-1/3$ in the $\Lambda$ rest frame.  A complication
arose for us because we measured the proton angular distribution in
the rest frame of the parent $\Sigma^0$.  This led to a value of $\nu
= -1/3.90$, as discussed in Appendix~\ref{app:sigmadecay}.  For the
$K^+\Lambda$ analysis $\nu=+1.0$.  Extraction of $P_{Y i}$ follows from
fitting the linear relationship of $I_i(\cos \theta_i)$ vs.~$\cos
\theta_i$.

The components of the measured hyperon polarization, $\vec{P}_Y$, are then
related to the polarization observables using the relations in
Eqs.~\ref{eq:obsx},~\ref{eq:obsp}, and~\ref{eq:obsz}.  The crucial
experimental aspect is that when the beam helicity is
reversed ($P_\odot \to -P_\odot$), so are the in-plane components of
the hyperon polarization.

In each bin of kaon angle  $\cos \theta_{K^+}^{c.m.} $, total system
energy $W$, and proton angle $\cos \theta_i $, let $N_\pm$ events 
be detected for a positive (negative) beam helicity according to 
\begin{equation}
N_\pm(\cos\theta_i) = \epsilon_K\epsilon_pQ_\pm\left[S I_i(\cos \theta_i) 
+ N_{BG}\right].
\label{eq:yield}
\end{equation}
$Q_\pm$ represents the number of photons with net helicity $\pm
P_\odot$ incident on the target. $S$ designates all cross section and
target related factors for producing events in the given kinematic
bin.  The spectrometer has a bin-dependent kaon acceptance defined as
$\epsilon_K$.  The protons from hyperon decay distributed according to
Eq.~\ref{eq:decay} are detected in bins, usually 10 in number, that
each have an associated spectrometer acceptance defined as
$\epsilon_p$.  In fact, $\epsilon_K$ and $\epsilon_p$ are correlated
since the reaction kinematics connect the places in the detector these
particles will appear.  This correlation is a function of $W$, $\cos
\theta_{K^+}^{c.m.}$, and $\cos \theta_i$, but is assumed to be beam
helicity independent.  We denote the correlated acceptance as
$\epsilon_K \epsilon_p$. The method used here avoids explicitly
computing this correlation. The term $N_{BG}$ designates events due to
``backgrounds'' from other physics reactions or from event
misidentifications.  The hyperon yield-fitting procedure discussed in
Sec.~\ref{sec:binningyield} removes $N_{BG}$, and the associated
residual uncertainty is discussed in Sec.~\ref{sec:systematics}.

If the beam helicity, $P_\odot$, can be ``flipped'' quickly and often,
then by far the most straightforward way to obtain the $C_i$ values is
to construct the ensuing asymmetry, $A$, as a function of proton
angle. In each proton angle bin we record the number of events,
$N_\pm$, in each beam helicity state and compute the corresponding
asymmetry as:
\begin{equation}
A(\cos\theta_i) = \frac{N_+ - N_-}{N_+ + N_-} =
\alpha\nu P_\odot C_i \cos\theta_i. 
\label{eq:asym}
\end{equation}

In this ratio the correlated detector acceptances and various
systematic effects cancel.  An exception would be if there were a
change in the track reconstruction efficiency due to a difference in
the beam intensity between the two beam polarization states.
Estimates of such phenomena proved negligibly small on the scale of
the results presented later.  If the beam intensity in the two beam
polarization states were not equal there would be a measured beam
intensity asymmetry (BIA) given by
\begin{equation}
A_{BIA} = \frac{Q_+ - Q_-}{Q_+ + Q_-}.
\label{eq:bia}
\end{equation}
This quantity is angle independent and therefore does
not influence the value of the slope of $A(\cos \theta_i)$.

\subsection{\label{sec:axis} Frame Transformation}

The hyperon polarizations were evaluated in the hyperon rest frames
according to the discussion in the previous sub-section.  The overall
center of mass (c.m.) frame of the reaction is reached by a boost
along the $\hat {z^\prime}$ axis, and we need to understand if and how
the polarization of the hyperons is changed in this
transformation. When boosting a baryon's spin projections from one
frame to another, one must take into account the Wigner-Thomas
precession that arises from the non-commutativity of rotations and
boosts.  In an initial frame $S$, suppose a particle has velocity
$\vec \beta $ $(=\vec p c/E)$ with respect to the boost direction at a
polar angle $\theta$.  In an arbitrary boosted frame $\tilde S$, let
the transformed velocity be described by $\tilde {\vec \beta}$ with
respect to the boost direction at a polar angle $\tilde\theta$.  Let
the corresponding boost parameters be $\Gamma$ for the frame boost,
$\gamma$ for the particle in the $S$ frame and $\tilde\gamma$ in the
boosted $\tilde S$ frame, where $\gamma = 1/\sqrt{1-\beta^2}$.  It can
be shown~\cite{giebink}~\cite{wignernote} that for an arbitrary boost
in the $\{\hat x , \hat z\}$ plane the Wigner-Thomas precession angle,
$\alpha_W$, about the $\hat y$ axis is given by
\begin{equation}
\sin\alpha_W = \frac{1+\Gamma}{\gamma + \tilde\gamma}
\sin(\theta - \tilde\theta).
\end{equation}
This relativistic rotation of the polarization direction is important,
for example, when transforming the laboratory-measured $(S)$ proton
recoil polarization in the reaction $p(\vec{e},e^\prime \vec{p})\pi^0$
to the center of mass frame of the virtual photon and target nucleon
$(\tilde S)$~\cite{wijesooriya}~\cite{schmieden}.  In this example the
boost direction is generally not collinear with the nucleon momentum in
$S$ or $\tilde S$, and the Wigner-Thomas precession angle can become
large.

In the present measurement, the boost to be performed is from the
hyperon rest frame $(S)$ to the c.m. frame of the real photon and
nucleon $(\tilde S)$.  Implicit in this discussion is that the
polarization is described in both frames with respect to the same
coordinate system. The boost is along the hyperon momentum direction,
so both $\theta$ and $\tilde\theta$ are zero.  Therefore the spin
precession angle $\alpha_W$ is identically zero for all hyperon
production angles.  The center-of-mass value for the hyperon
polarization is thus the same as it is in the hyperon rest frame.  We
must measure $\vec P_Y$ in the hyperon rest frame, but it is the same
in the overall reaction c.m. frame.

\section{\label{sec:setup}EXPERIMENTAL SETUP}

The data analyzed to measure $C_x$ and $C_z$ were recorded by the CLAS
spectrometer in Hall B at Jefferson Lab.  Data were produced at two
different electron energies, $E_{elec} = 2.4$ and $2.9$ GeV.  The 2.4
GeV data set was previously analyzed in combination with a third data
set at 3.1 GeV to extract differential cross sections
\cite{bradforddsdo} and for $\Lambda$ and $\Sigma^0$ recoil
polarizations \cite{mcnabb}.  These present measurements are the first
reported results from the 2.9 GeV data set.  All data sets were
recorded under the same (``g1c'') run conditions.  In the previous
papers, the beam polarization and measurement of the in-plane recoil
polarization were not relevant, but now we discuss these points.

The incident polarized electron beam was used to create a secondary
beam of circularly polarized photons using the Hall B photon tagging
system.  Bremsstrahlung photons were produced by colliding the
longitudinally polarized electron beam with a gold foil radiator.  The
residual momenta of the recoiling electrons were measured with a
hodoscope behind a dipole magnetic field.  This information was used
to determine the energy and predict the arrival time of photons
striking the physics target.  The energy range of the tagging system
spanned from 20\% to 95\% of the endpoint energy.  The rate of tagged
photons was about $1.4 \times 10^7$/sec. Detailed information about
the CLAS photon tagging system is given in \cite{tagger}.  The physics
target consisted of a 18 cm long cell of liquid hydrogen located at
the center of the CLAS detector.

The CLAS detector is a multi-particle large acceptance spectrometer
that incorporates a number of subsystems.  The start counter (SC), a
scintillator counter surrounding the target, was used to obtain a fast
timing signal as particles left the target. The tracking system
of the detector included 34 layers of drift chamber cells.  A toroidal
magnetic field provided by a superconducting magnetic bent the
trajectories of charged particles through the tracking volume for
momentum determination.  For this experiment, the magnetic field was
operated so that positively charged particles were bent outward, away
from the beamline.  Finally, as particles left the detector, an outer
scintillator layer, the time-of-flight (or TOF) array made a final
timing measurement.  The readout trigger required coincidence between
timing signals from the photon tagger, SC, and the TOF.  More general
information about the detector and its performance can be obtained
from Ref.~\cite{clas0}; the detector configuration at the time of this
experiment is further detailed in Refs.\cite{mcnabbthesis,
bradfordthesis}.

\subsection{\label{sec:beam} Beam Polarization}

Extraction of $C_x$ and $C_z$ from the beam helicity asymmetry, as
discussed in Section \ref{sec:analysis}, required accurate knowledge
of the photon beam polarization.  Since Hall B has no Compton
polarimeter to directly measure the photon beam polarization, this
information was obtained through a two step process. The polarization
of the incident electron beam was measured with a M{\o}ller
polarimeter, and a well-known formula then gave the polarization of
the secondary photon beam.

The Hall B M{\o}ller polarimeter~\cite{moeller} is a dual-arm
coincidence device which exploits the helicity dependence of M{\o}ller
scattering to measure the polarization of the incident electron beam.
Beam electrons were scattered elastically from electrons in the
polarimeter target.  A pair of quadrupole magnets collected the
scattered electrons on a pair of scintillation counters.
Helicity-dependent yields, $N_+$ and $N_-$, were recorded.  From these
yields, the electron beam polarization was measured according to
\begin{equation}
A_{elec} =\frac{N_+ - N_-}{N_++N_-}=A_zP_{elec}P_T,
\end{equation}
where $A_{elec}$ is the helicity-dependent asymmetry, $A_z$ is the
analyzing power of the polarimeter iron foil target, $P_T$ is the
polarization of the target material, and $P_{elec}$ is the
polarization of the incident beam.

Operation of the M{\o}ller polarimeter disrupted the beam and was
periodically done separately from the main data taking.  The various
measurements were averaged for each run period and reported as a
single polarization.  The results are shown in
Table~\ref{epolarization}.  The uncertainties shown are estimated
random and averaging uncertainties.  The estimated systematic
uncertainty on the M{\o}ller measurements was $\pm
3\%$~\cite{moeller}.  The values of $P_{elec}$ are typical of the
Jefferson Lab electron beam when using a strained GaAs cathode and
laser to produce electrons.

\begin{table}
\begin{center}
\begin{tabular}{|c|c|}
\hline
Beam Energy & $e^-$ Beam Polarization\\
\hline
 2.4 GeV & $0.654 \pm 0.015$\\ 2.9 GeV & $0.641 \pm 0.012$\\
\hline
\end{tabular}
\end{center}
\caption{Electron beam polarizations, $P_{elec}$, used in these measurements.}
\label{epolarization}
\end{table}

The polarization of the beam was flipped at the injector to the
accelerator at a rate of 30 Hz in a simple non-random
$+-+-$... sequence.  The beam helicity state was recorded event by
event in the data stream.  

The energy-dependent circular polarization, $P_\odot(E_\gamma)$, of
the photons originating from the bremsstrahlung of the longitudinally
polarized electrons on a radiator was computed using the expression
\begin{equation}
P_\odot(E_\gamma) = \frac{y(4-y)}{4-4y+3y^2} P_{elec},
\end{equation}
where $y= E_\gamma/E_{elec}$ is the fraction of photon energy
$E_\gamma$ to beam energy $E_{elec}$, and $P_{elec}$ is the
polarization of the electron beam.  This expression is a slightly
rewritten version of Eq. 8.11  in Ref.~\cite{maximon}.  The photon
polarization is maximal at the bremsstrahlung endpoint and falls
rather slowly with decreasing photon energy.  Over the photon energy
range used in this measurement we had $0.440<P_\odot/P_{elec}<0.995$.

\section{\label{sec:analysis}DATA ANALYSIS}

\subsection{\label{sec:pid}Particle Identification and Event Selection}

Particle identification for this analysis was identical to that
reported for our differential cross section
analysis~\cite{bradforddsdo}.  In general, particle identification was
based on time-of-flight.  For each track of momentum $\vec{p}$, we
compared the measured time-of-flight, $TOF_m$, to a hadron's expected
time-of-flight, $TOF_h$, for a kaon, pion, or proton of identical
momentum.  Cuts were placed on the difference between the measured and
expected time-of-flight, $\Delta TOF = TOF_m - TOF_h$.

Because our measurement technique relied on the self-analyzing nature
of the hyperon recoil polarizations, we selected events exclusively
involving the charged final state of the decaying hyperons according
to $ \Lambda \to p \pi^- $ and $\Sigma^0 \to \gamma \Lambda \to \gamma
p \pi^-$.  Three criteria were used to select such events.  First, all
events were required to have both a $K^+$ and a proton track.  Second,
events were required to have a $p\left(\gamma,K^+ \right)Y$ missing
mass consistent with the mass of a $\Lambda$ or $\Sigma ^0$
hyperon. Finally, we did not require explicit detection of the $\pi^-$
from the hyperon decays, but we required that the $p\left( \gamma,
K^+p\right) \pi^- \left( \gamma \right)$ missing mass be consistent
with a $\pi^-$ (or $\gamma \pi^-$ for $K^+ \Sigma^0$ events).  While
CLAS was able to detect some of the $\pi^-$ tracks directly,
acceptance losses reduced the event statistics excessively.  To
further increase the acceptance of events, we relaxed the fiducial
cuts employed in the cross section analysis to permit more tracks near
the detector edges.  This increased the yield of useful events by a
factor of about 60\%.  Specific cuts to select each hyperon species
were developed and are detailed in Ref.~\cite{bradfordthesis}.

\subsection{\label{sec:binningyield}Binning and Yield Extraction}

Hyperon yields were divided into kinematic bins in photon energy
($E_{\gamma}$), recoiling kaon angle in the c.m.~frame $(\cos
\theta_K^{c.m.})$, the angle of the decay proton in the hyperon rest
frame $(\cos \theta_i )$, and the helicity of the incident photon beam.
Bin widths and limits are detailed in Table~\ref{bintable}.

\begin{table}
\begin{center}
\begin{tabular}{|c|c|c|c|c|c|c|c|c|c|}
\hline & 
\multicolumn{3}{|c|}{$E_{\gamma}$ (GeV)}         &
\multicolumn{3}{|c|}{$\cos \theta_{K^+}^{c.m.}$} &
\multicolumn{3}{|c|}{$\cos \theta_{i}$}          \\
\hline Channel & Low & High & $\Delta$ & Low & High & $\Delta$ & Low &
High & $\Delta$\\ 
\hline $K^+\Lambda$ & 0.9375 & 2.7375 &
0.1 & -0.85 & 0.95 & 0.2 & -1.0 & 1.0 & 0.2 \\ 
\hline $K^+\Sigma^0$ 
& 1.1375 & 2.7375 & 0.1 & -0.85 & 0.95 & 0.3 & -1.0 & 1.0 &
0.4 \\ \hline
\end{tabular}
\end{center}
\caption{Binning for $C_{x}$ and $C_{z}$.  $K^+ \Lambda$ observables
used a total of 3420 bins, while the $K^+ \Sigma ^0$ observables had
1020 bins.  The ``low'' and ``high'' of each kinematic variable in the
table are the edges of our kinematic coverage, not the bin
centers. The $\Delta$ columns give the width of the bins.}
\label{bintable}
\end{table}

Two independent hyperon yield extractions were performed in each bin.
The first extraction employed a fit to the $p \left( \gamma,
K^+\right) Y$ missing mass spectrum in the region of the $\Lambda$ and
$\Sigma^0$ mass peaks (1.0 to 1.3 GeV/c$^2$).  Hyperon peaks were each
fit to a Gaussian line shape, while the backgrounds were modeled with a
polynomial of up to second order.  Since the background shape varied
slowly across the kinematic coverage, the background shape employed in
the fits was selected on a bin-by-bin basis; see
Ref.~\cite{bradforddsdo} for sample yield fits.  The second extraction
method relied on side-band subtraction in which the background was
assumed to be smooth under the hyperons.

\subsection{\label{sec:asymmetry}Asymmetry Calculation and Slope Extraction}

Within each $\{ E_\gamma$ , $\cos \theta_{K^+}^{c.m.}$, $\cos\theta
_i\}$ bin, the helicity dependent yields were used to calculate the
beam-helicity asymmetry according to the sum of Eqs.~\ref{eq:asym} and
~\ref{eq:bia}.  Two different versions of this asymmetry were
calculated.  The fit-based asymmetry method, or FBA, was largely based
on yields determined by the Gaussian-plus-background fits, with the
side-band yields used in bins where the fits failed.  The second
calculation employed only sideband-subtracted asymmetries, or SBA; all
fits were turned off for this calculation.

\begin{figure}
\vspace{1.0cm}
\resizebox{0.5\textwidth}{!}{\includegraphics{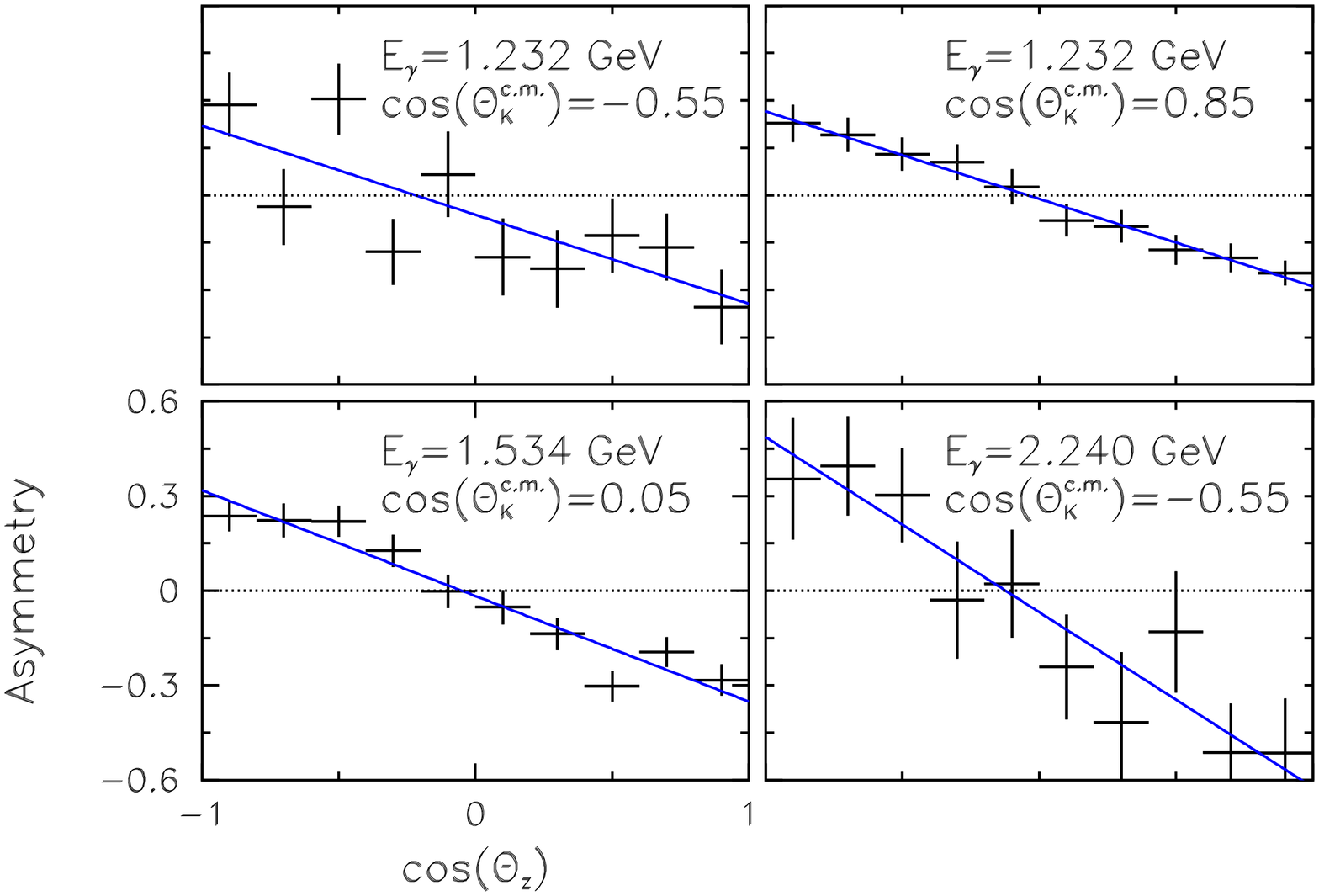}}
\caption{(Color online) Representative hyperon yield asymmetries as a
  function of proton decay angle for the case of the $C_z$ observable
  for the $\Lambda$. The scales are the same in all plots. }
\label{fig:adistLambda}       
\end{figure}

\begin{figure}
\vspace{1.0cm}
\resizebox{0.5\textwidth}{!}{\includegraphics{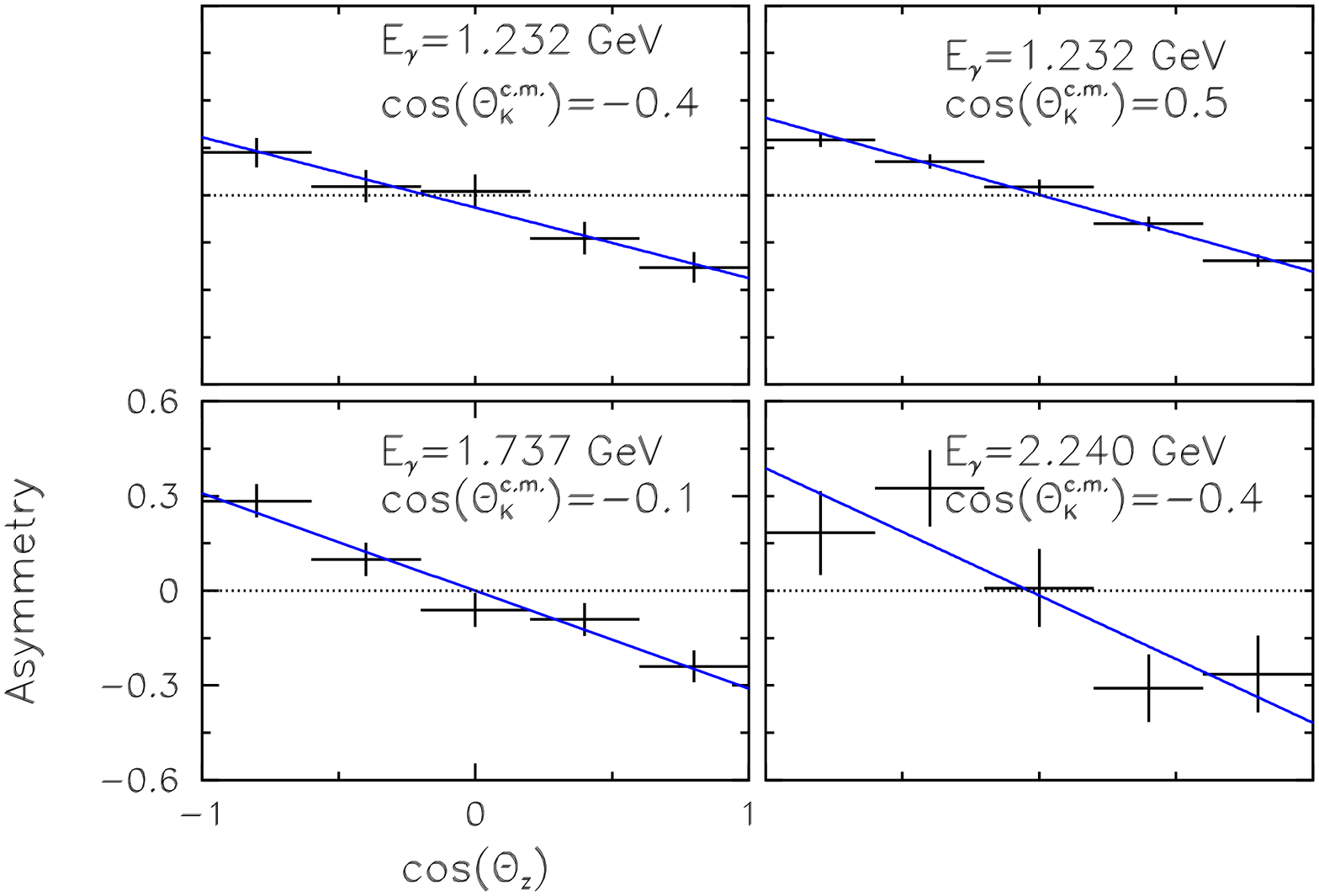}}
\caption{(Color online) Representative hyperon yield asymmetries as a
  function of proton decay angle for the case of the $C_z$ observable
  for the $\Sigma^0$. The scales are the same in all plots. }
\label{fig:adistSigma0}       
\end{figure}

The asymmetries were computed vs.~$\cos \theta _i$, and linear fits
were used to extract the slopes of the distributions.  The free
parameters were the product $\alpha\nu P_{\odot}C_i$ and $A_{BIA}$ in
Eq~\ref{eq:bia}.  Some sample distributions are shown in
Figs.~\ref{fig:adistLambda} and \ref{fig:adistSigma0} for the
$\Lambda$ and $\Sigma^0$ cases, respectively.  In general, the
asymmetry distributions were very well fit with a sloped line.
Counting statistics were poorest at lower photon energies and backward
kaon angles, where the cross sections were smallest and the kaon decay
probability was largest, but the statistics improved rapidly for mid-
to forward-going kaons and higher photon energies.  Results with and
without constraining $A_{BIA}$ to be zero were in very good agreement,
but we did not constrain this offset to be zero to avoid bias from
this source. The average fitted value was $A_{BIA} = 0.002$ with a
standard deviation of 0.027.

\begin{figure}
\vspace{1.0cm}
\resizebox{0.4\textwidth}{!}{\includegraphics{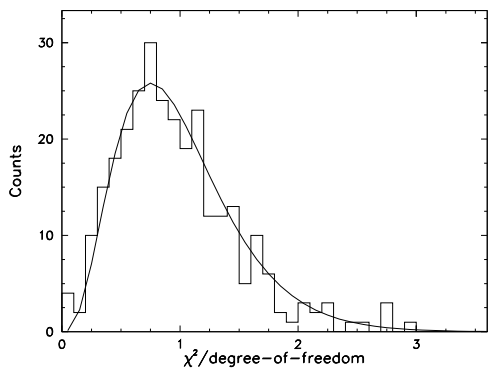}}
\caption{(Color online) Distribution of the reduced $\chi^2$ values
for fits with 8 degrees of freedom for the $C_z$ fits in the
$K^+\Lambda$ case.}
\label{fig:chi2}       
\end{figure}

The overall fit quality is well summarized by the distribution of
$\chi ^2$ per degree-of-freedom.  Figure~\ref{fig:chi2} shows this
distribution for the linear fits used in the measurement of $C_z$ for
the $K^+ \Lambda$ case.  This figure shows that the actual $\chi^2$
distribution is consistent with the expected distribution, indicated
by the smooth curve superimposed on the histogram.  The actual and
expected $\chi^2$ distributions were consistent for all results
reported in this paper.

\begin{figure}
\vspace{1.0cm}
\resizebox{0.5\textwidth}{!}{\includegraphics{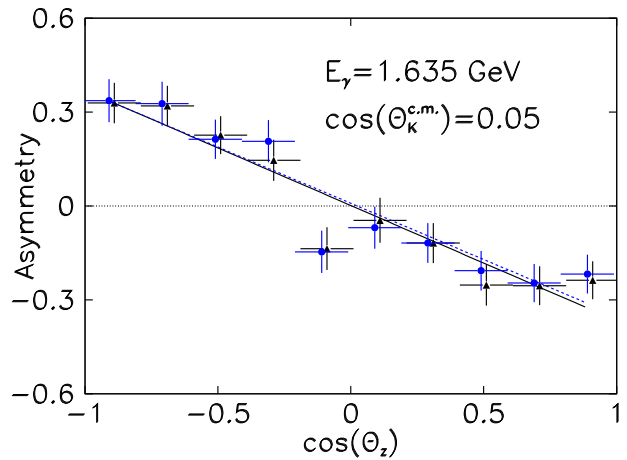}}
\caption{(Color online) Hyperon yield asymmetries as a function of
  proton decay angle.  The two sets of points were obtained via the
  FBA (black triangles) and SBA (blue circles) methods,
  respectively. The two fitted lines, which are proportional to $C_z$
  for the $K^+\Lambda$ case, show a visible difference, as discussed
  in the text.  }
\label{fig:2dist}       
\end{figure}

Within each $\{E_{\gamma}$, $\cos \theta _{K^+}^{c.m.} \}$ kinematic
bin, we compared the FBA and SBA asymmetries, as shown in
Fig.~\ref{fig:2dist}.  In the large majority of kinematic bins, the
distributions were statistically consistent, though in a few bins the
two methods differed significantly.  The final results were based on
the asymmetry calculation (FBA or SBA) that was fit best by the
straight line. The differences were used to estimate the systematic
uncertainty associated with the yield extraction.

\subsection{\label{sec:systematics}Systematic Uncertainties}

As shown in Eq.~\ref{eq:asym}, four factors are key to
measurements of $C_x$ and $C_z$: (1) the beam helicity asymmetry, (2)
the beam polarization, (3) the weak decay asymmetry parameter, and (4)
the dilution factor.  Uncertainties on each one of these factors may
contribute to systematic uncertainty in our results.

We studied dependence of the beam helicity asymmetry on the yield
extraction method.  As discussed in Section \ref{sec:asymmetry}, we
performed two different yield extractions and we calculated two
versions of the beam helicity asymmetry, the fit-based asymmetry (FBA)
and the side-band subtracted asymmetry (SBA).  Within each $\{
E_{\gamma}$, $\cos \theta _K ^{c.m.} \}$ kinematic bin, we fit each
asymmetry distribution independently and measured the difference
between the extracted slopes.  This slope difference was interpreted
as a point-to-point systematic error due to the yield extraction
method.  This slope difference was added in quadrature with the error
on the extracted slope and propagated through the analysis.  The good
agreement between the methods was the basis for our treating $N_{BG}$
in Eq.~\ref{eq:yield} as negligible.  Uncertainties in this paper, then,
include statistical errors plus the estimated point-by-point
systematic error due to the yield extraction.

The CLAS M{\o}ller polarimeter has uncertainties in the analyzing
power of the reaction and in the polarimeter's target
polarization~\cite{moeller}, which resulted in a systematic
uncertainty of $\pm 0.016$ on the final observables.  Measurements
from the polarimeter also had their own statistical uncertainties,
shown in Table~\ref{epolarization}, which also contributed to the
global systematic error. When propagated, the contribution to the
systematic error is $\pm 0.022$.

The $\Lambda$ weak decay asymmetry parameter $\alpha$ has a
well-documented uncertainty~\cite{pdg06} of $\pm 0.013$.  The
contribution to the global systematic uncertainty is then $\pm 0.020$.
The dilution factor $\nu$, discussed in Appendix~\ref{app:sigmadecay},
is a purely computational quantity that is assumed to have negligible
uncertainty compared to the other sources discussed here.

Our analysis method for $C_x$ and $C_z$ should result in a vanishing
measured transverse polarization of the hyperons, $P_{Y y}$. That is,
the helicity asymmetry of the out-of-plane projection of the hyperons'
polarization, as defined in Fig.~\ref{fig:axes}, must be zero.  This
test formed a useful systematic check of our method.  To measure
``$C_y$'', the same analysis procedure was applied as for $C_x$ and
$C_z$, the only difference being that the proton direction was
projected onto $\hat{y}$ in the hyperon rest frame.  The results were
consistent with zero over a large range of kinematics, but $C_y$ was
statistically nonzero for fairly forward kaon c.m. angles for both
hyperons.  This was attributed to the measurement $\hat y = \hat
\gamma \times \hat K$ being less accurate at very forward kaon
laboratory angles due to detector geometry and resolution effects.
Such distortions would similarly affect $C_x$, for example by letting
a large $P_{Y z}$ mix into small values of $P_{Y x}$.  As a result,
there is an angle-dependent systematic uncertainty of $\pm 0.08$ for
$\Lambda$ observables at $\cos \theta_{K^+}^{c.m.} > 0.55$, and $\pm
0.17$ for $\Sigma^0$ observables at $\cos \theta _{K^+}^{c.m.}  >
0.35$.

When summed in quadrature, 
we estimate a total global systematic
uncertainty for the $K^+\Lambda$ results as 
$\pm 0.03$ for $\cos \theta^{c.m.}_{K^+} < 0.55$ and
$\pm 0.09$ for $\cos \theta^{c.m.}_{K^+} > 0.55$.  
We estimate a total global systematic
uncertainty for the $K^+\Sigma^0$ results as 
$\pm 0.03$ for $\cos \theta^{c.m.}_{K^+} < 0.35$ and
$\pm 0.17$ for $\cos \theta^{c.m.}_{K^+} > 0.35$.  
The systematic uncertainty in $W$ was
$\pm 2$ MeV at the bin centers.

\section{\label{sec:results}RESULTS}

\subsection{\label{sec:cxczdata}$C_x$ and $C_z$ Results for $K^+ \Lambda$}

\begin{figure*}[ht]
\vspace{1.0cm}
\resizebox{0.8\textwidth}{!}{\includegraphics[70,50][600,450]{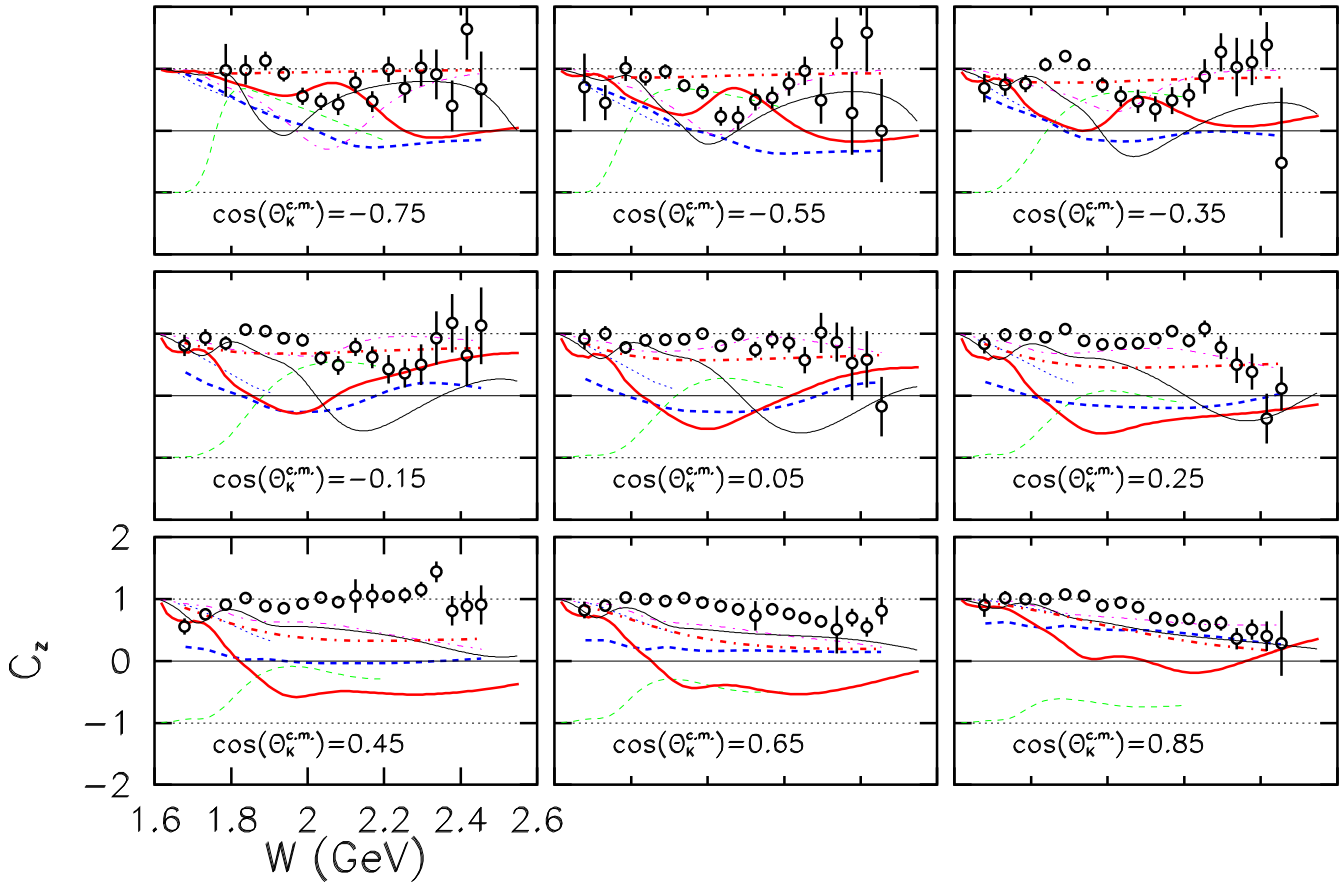}}
\caption{(Color online) The observable $C_z$ for the reaction
$\vec\gamma + p \to K^+ + \vec\Lambda$, plotted as a function of the
c.m.~energy $W$.  The circles are the results of this measurement,
with uncertainties discussed in the text.  
The thin dashed (green) curves are from Kaon-MAID~\cite{maid}, 
the thick solid (red) curves are from SAP~\cite{saghai2}, 
the thick dashed (blue) curves are from BG~\cite{sarantsev2},  
the thin solid (black) curves are from RPR~\cite{corthals}, and
the thick dot-dashed (magenta) curves are from GENT~\cite{jan}.  
The thick dot-dashed (red) curves are from GLV~\cite{lag1,lag2}.  
The thin dotted (blue) curves are from SLM~\cite{shklyar}.  
}
\label{fig:czlambda_w}       
\end{figure*}

\begin{figure*}[ht]
\vspace{1.0cm}
\resizebox{0.8\textwidth}{!}{\includegraphics[70,50][600,450]{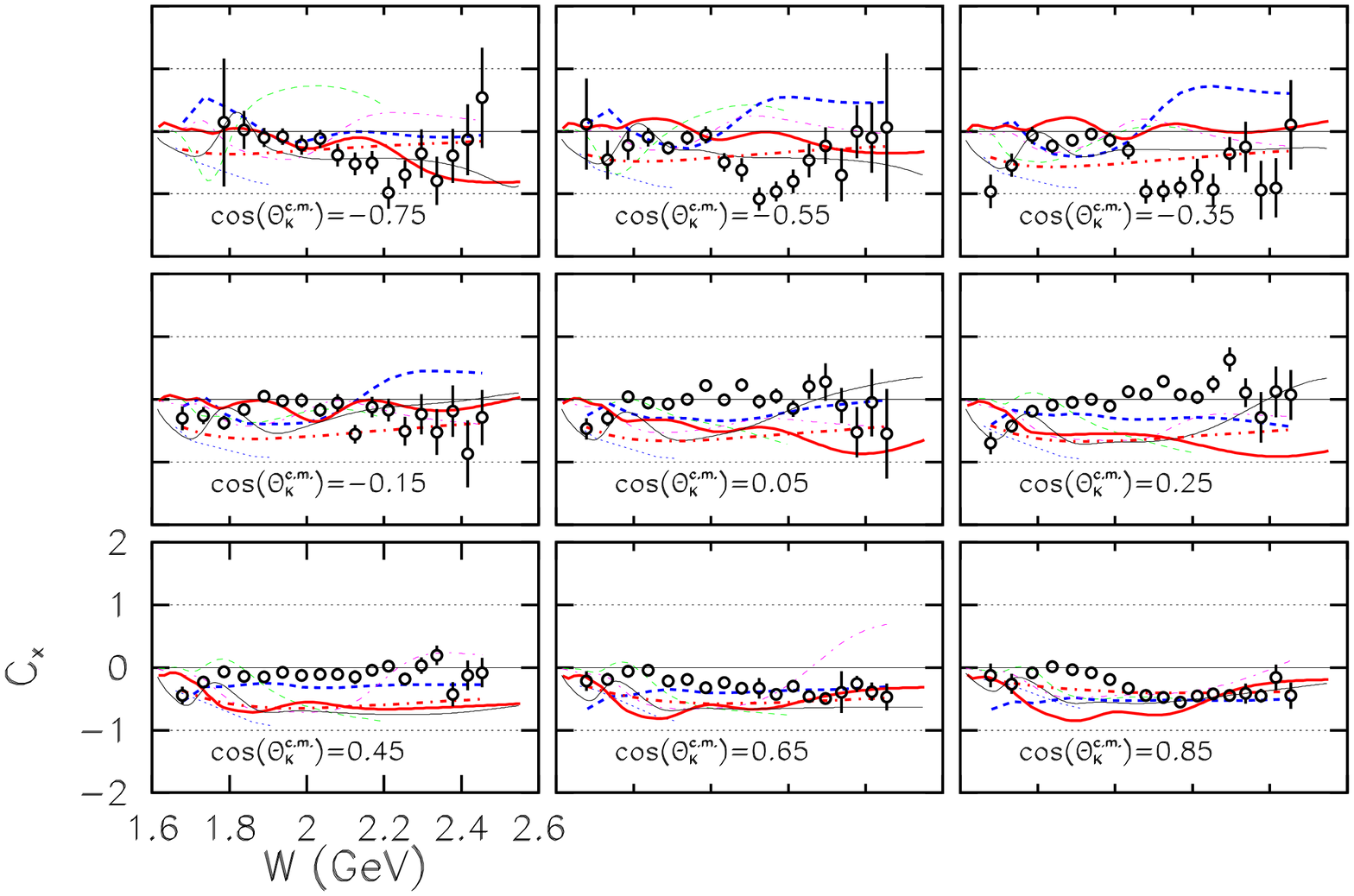}}
\caption{(Color online) The observable $C_x$ for the reaction
$\vec\gamma + p \to K^+ + \vec\Lambda$, plotted as a function of the
c.m.~energy $W$.  The circles are the results of this measurement,
with uncertainties discussed in the text.  
The thin dashed (green) curves are from Kaon-MAID~\cite{maid}, 
the thick solid (red) curves are from SAP~\cite{saghai2}, 
the thick dashed (blue) curves are from BG~\cite{sarantsev2},  
the thin solid (black) curves are from RPR~\cite{corthals}, and
the thick dot-dashed (magenta) curves are from GENT~\cite{jan}.  
The thick dot-dashed (red) curves are from GLV~\cite{lag1,lag2}.  
The thin dotted (blue) curves are from SLM~\cite{shklyar}.  
}
\label{fig:cxlambda_w}       
\end{figure*}

\begin{figure*}[ht]
\resizebox{0.6\textwidth}{!}{\includegraphics[70,50][600,500]{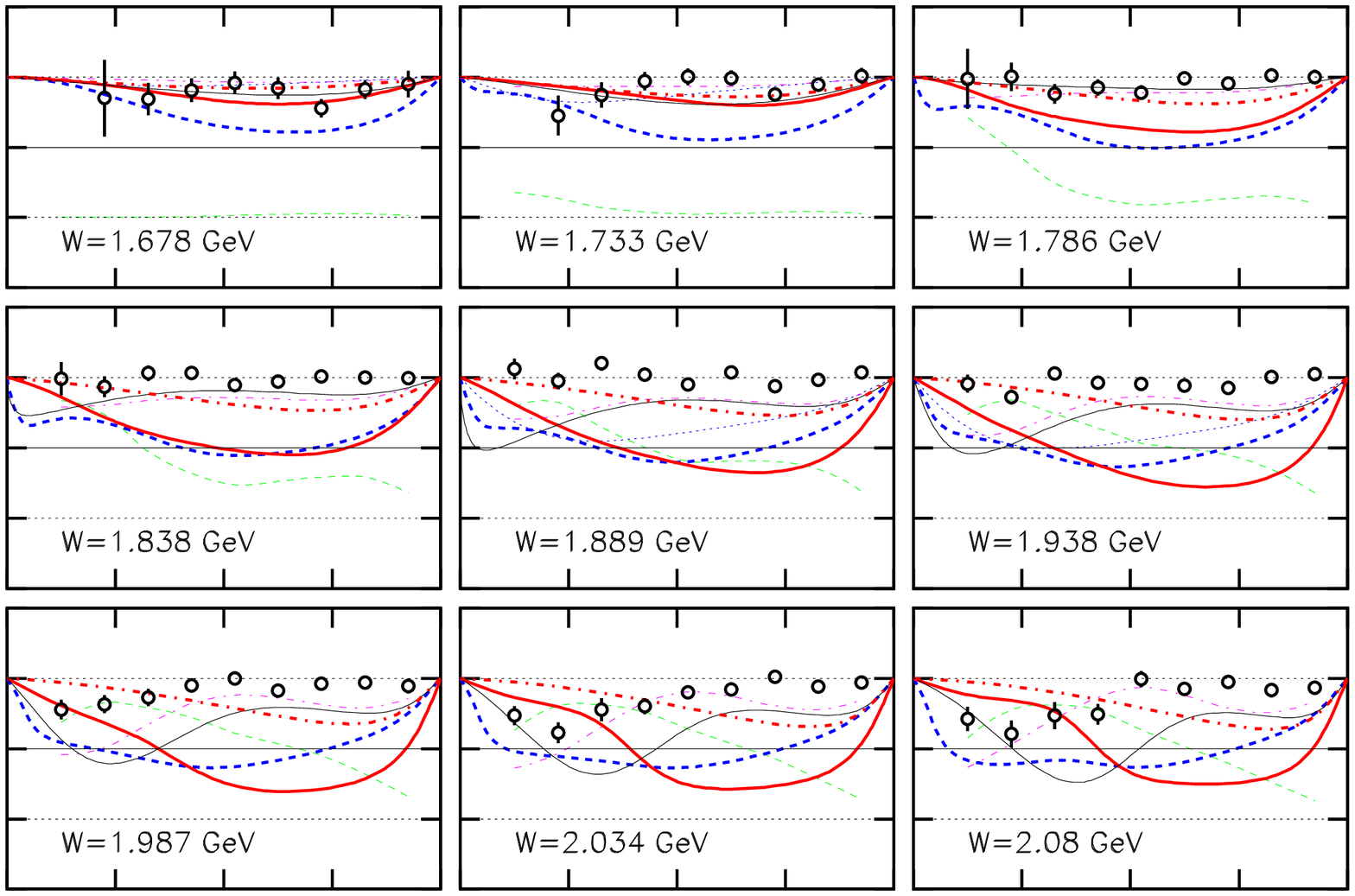}}
\resizebox{0.6\textwidth}{!}{\includegraphics[70,50][600,440]{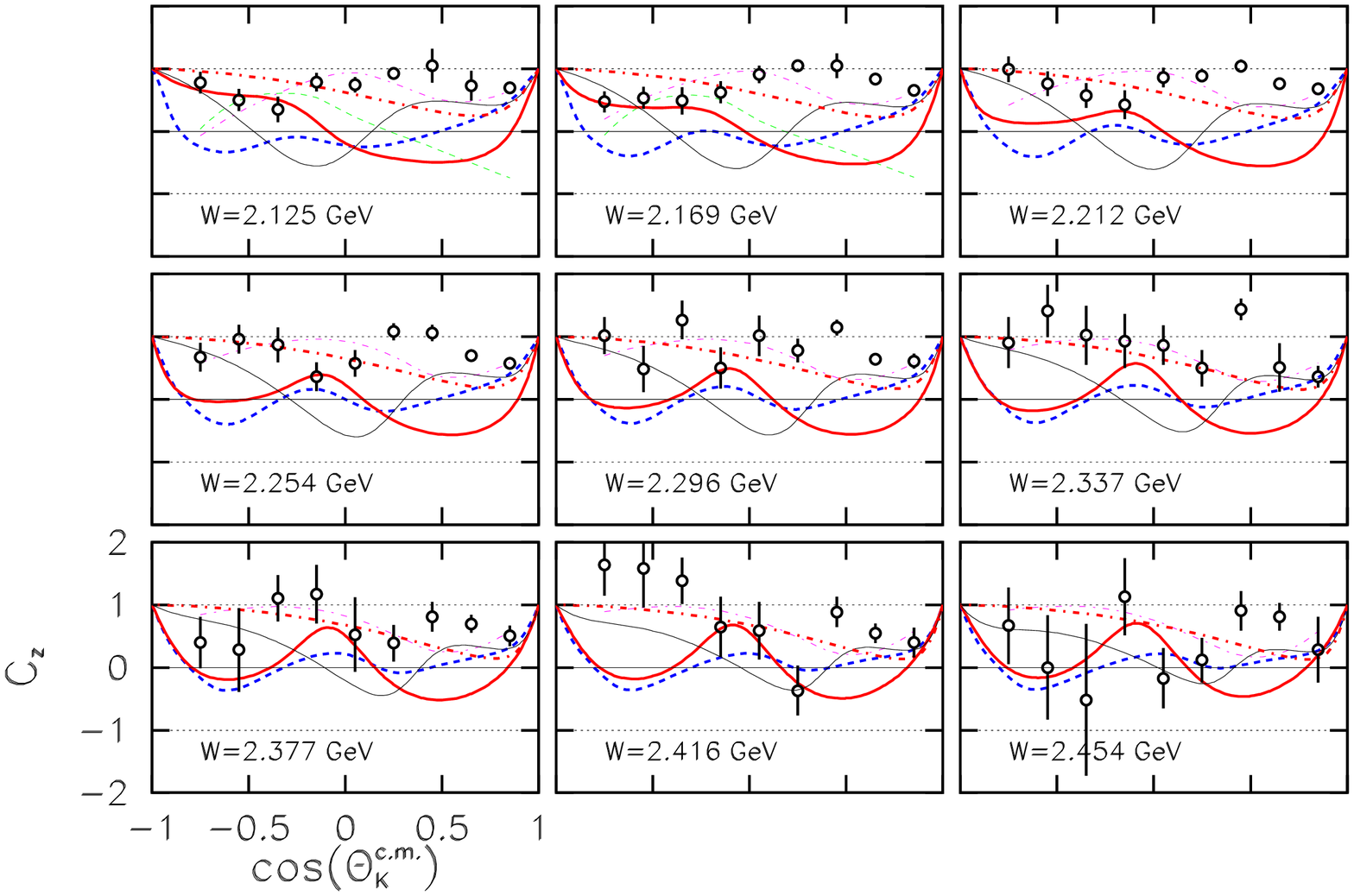}}
\caption{(Color online) The observable $C_z$ for the reaction
$\vec\gamma + p \to K^+ + \vec\Lambda$, plotted as a function of the
kaon angle.  The 18 panels are for increasing values of $W$ in steps
of about 50 MeV.  The circles are the results of this measurement,
with uncertainties discussed in the text.  
The thin dashed (green) curves are from Kaon-MAID~\cite{maid}, 
the thick solid (red) curves are from SAP~\cite{saghai2}, 
the thick dashed (blue) curves are from BG~\cite{sarantsev2},  
the thin solid (black) curves are from RPR~\cite{corthals}, and
the thick dot-dashed (magenta) curves are from GENT~\cite{jan}.  
The thick dot-dashed (red) curves are from GLV~\cite{lag1,lag2}.  
The thin dotted (blue) curves are from SLM~\cite{shklyar}.  
}
\label{fig:czlambda_angle}       
\end{figure*}

\begin{figure*}[ht]
\resizebox{0.6\textwidth}{!}{\includegraphics[70,50][600,500]{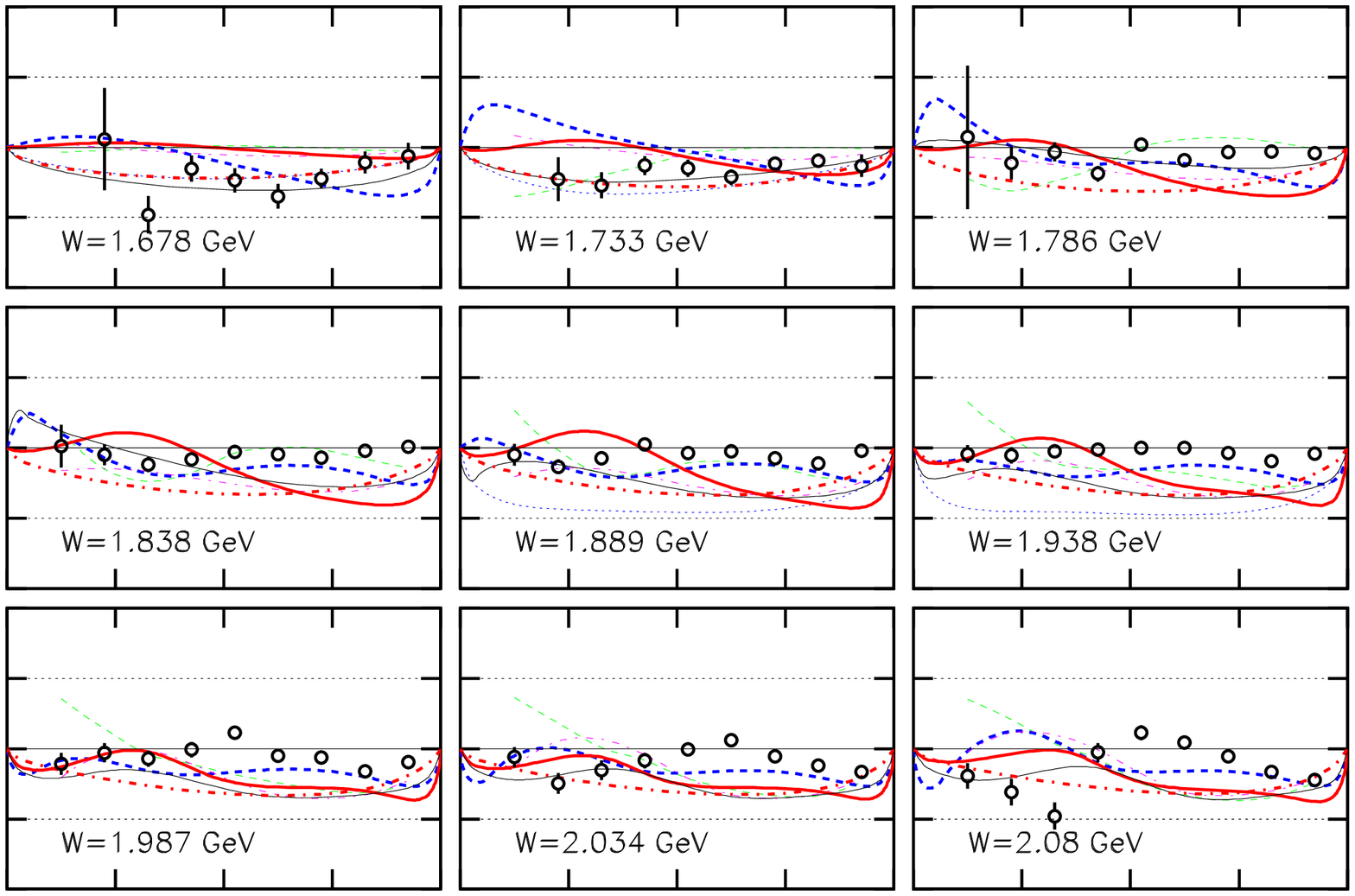}}
\resizebox{0.6\textwidth}{!}{\includegraphics[70,50][600,440]{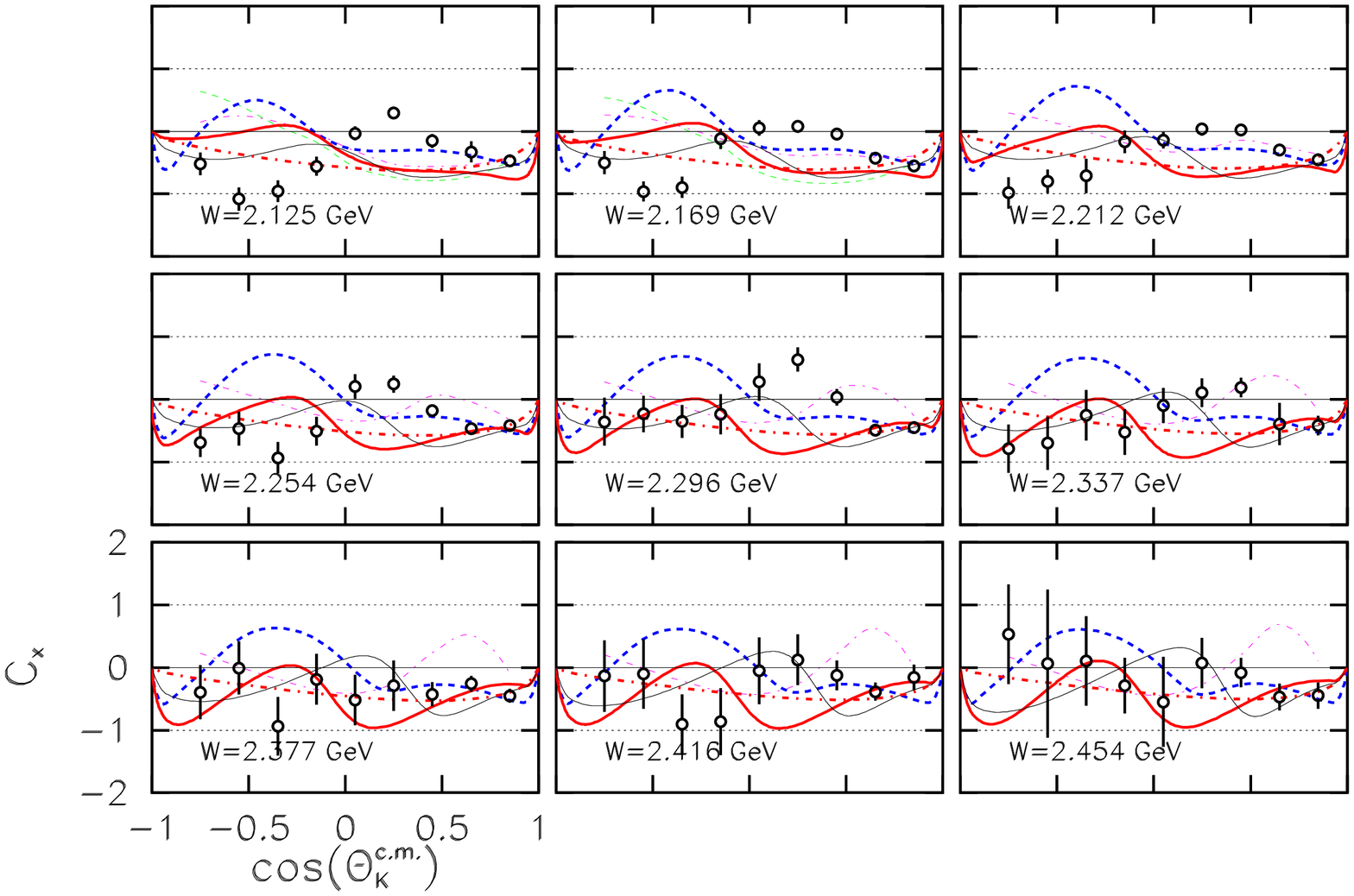}}
\caption{(Color online) The observable $C_x$ for the reaction
$\vec\gamma + p \to K^+ + \vec\Lambda$, plotted as a function of the
kaon angle.  The 18 panels are for increasing values of $W$ in steps
of about 50 MeV.  The circles are the results of this measurement,
with uncertainties discussed in the text.  
The thin dashed (green) curves are from Kaon-MAID~\cite{maid}, 
the thick solid (red) curves are from SAP~\cite{saghai2}, 
the thick dashed (blue) curves are from BG~\cite{sarantsev2},  
the thin solid (black) curves are from RPR~\cite{corthals}, and
the thick dot-dashed (magenta) curves are from GENT~\cite{jan}.  
The thick dot-dashed (red) curves are from GLV~\cite{lag1,lag2}.  
The thin dotted (blue) curves are from SLM~\cite{shklyar}.  
}
\label{fig:cxlambda_angle}       
\end{figure*}

As discussed in Section~\ref{sec:method}, the transfer of circular
polarization from the incident photon beam to the recoiling hyperons
leads to the observable $C_z$ along the beam direction and $C_x$ in
the $\hat\gamma \times \hat K$ reaction plane and perpendicular to the
beam axis. The results for the $W$ dependence for the reaction
$\vec\gamma+p\to K^+ + \vec\Lambda$ are given in
Figs.~\ref{fig:czlambda_w} and~\ref{fig:cxlambda_w}.  The same results
are presented as a function of kaon c.m.~angle in
Figs.~\ref{fig:czlambda_angle} and~\ref{fig:cxlambda_angle}.  The
given error bars combine the statistical uncertainties and the
estimated point-to-point systematic uncertainties arising from the
fits to the helicity asymmetries.


It is immediately evident in these results that, qualitatively, the
photon polarization is largely transferred to the $\Lambda$ hyperon
along the $\hat z$ direction in the c.m. frame.  Figure
~\ref{fig:czlambda_angle} shows that from threshold up to about 1.9
GeV the $\Lambda$ data exhibit $C_z\sim +1$, which means it has nearly
the full polarization transferred to it, irrespective of the
production angle of the kaon.  For higher values of $W$ one can see
fall-offs of the value of $C_z$ as a function of kaon c.m. angle.
However, for kaons produced in the forward hemisphere, the nearly full
transfer effect is present up to about 2.1 GeV, as seen in
Fig.~\ref{fig:czlambda_w}.  Above this energy the forward-angle value
of $C_z$ decreases with increasing $W$.  The concomitant values of
$C_x$ are generally closer to zero, as seen in
Fig.~\ref{fig:cxlambda_w}, with significant excursions to negative
values for a combination of backward kaon angle and high energies, and
again for the very forward angles and higher energies.

This striking observation of large and quasi-constant values of $C_z$
is why we chose to present our results in the $\{x, z\}$ coordinate
system rather than the $\{x^\prime , z^\prime\}$ system.  It can be
interpreted in terms of a picture wherein the photon excites an
$s$-channel resonance which decays with no orbital angular momentum,
$L$, along the $\hat z$ direction.  In a simple classical picture of a
two-particle $s$-channel interaction, any orbital angular momentum is
normal to $\hat z$.  To conserve the $z$ component of angular
momentum, the hyperon must then carry it in the form of spin
polarization.  In the case of $K^+\Lambda$ near threshold, the
reaction is thought to be dominated by the $S_{11}$ partial wave, for
which this argument applies.  There is no reason for this picture to
hold up, however, when multiple amplitudes conspire to result in the
observed polarization.  Thus, it is surprising how ``simple'' the
result for $K^+\Lambda$ appears.

At higher energies and backward kaon c.m. angles the ``simple''
pictures gives way to more interference structure in both $C_z$ and
$C_x$.  For example, in Fig.~\ref{fig:cxlambda_w}, $C_x$ takes values
close to $-1.0$ for $\cos\theta_{K^+}^{c.m.} <-0.35$ and $W>2.1$ GeV.
Also at the most forward angles for $W>2.1$ GeV there is a monotonic
trend downward in both $C_z$ and $C_x$.

\subsection{\label{sec:pdata}Combining $C_x$, $C_z$ Results with Results for $P$}

There are several inequalities that must be satisfied by the
observables available in pseudo-scalar meson
photoproduction~\cite{barker, goldstein, tabakin1}.  
Artru, Richard, and
Soffer~\cite{soffer} pointed out 
that for a circularly polarized beam there is a rigorous inequality
\begin{equation}
R^2 \equiv P^2 + C_x^2 +C_z^2 \leq 1
\label{eq:are}
\end{equation}
among the three polarization observables, where $P$ is the same as the
measured $P_{Yy}$, the induced recoil polarization of the baryon.  For
a $100\%$ circularly polarized photon beam, $\vec R$ is equivalent to
$\vec P_Y$ defined in Eq.~\ref{eq:density}.  In this case the
relationship says that the magnitude of the three orthogonal
polarization components may have any value up to unity.  There is no
{\it a priori} requirement that the hyperon be produced fully
polarized except in the extreme forward and backward directions where
orbital angular momentum plays no role.  Any rotation of the
coordinate system about $\hat y$ would redefine the $C_i$ but leave
the inequality unchanged, since the baryon polarization transforms as
a 3-vector under spatial rotations.

A significant test of the present results for $C_x$ and $C_z$ is
therefore compatibility with the previously-published~\cite{mcnabb}
results for the induced hyperon recoil polarization $P$.  (We note
that those earlier data have been confirmed up to $E_\gamma=1.5$ GeV
by measurements at GRAAL~\cite{graal}.)  While the helicity
asymmetries used in the present measurement are sensitive to $C_x$ and
$C_z$, the $\hat y$ helicity asymmetry must be zero by reason of
parity conservation.  On the other hand, our previous measurement
ignored the beam polarization information and was sensitive to $P$ but
not $C_x$ and $C_z$.  Taken together, the measurements should obey the
constraint given above.

\begin{figure*}
\resizebox{1.0\textwidth}{!}{\includegraphics[10,50][700,500]{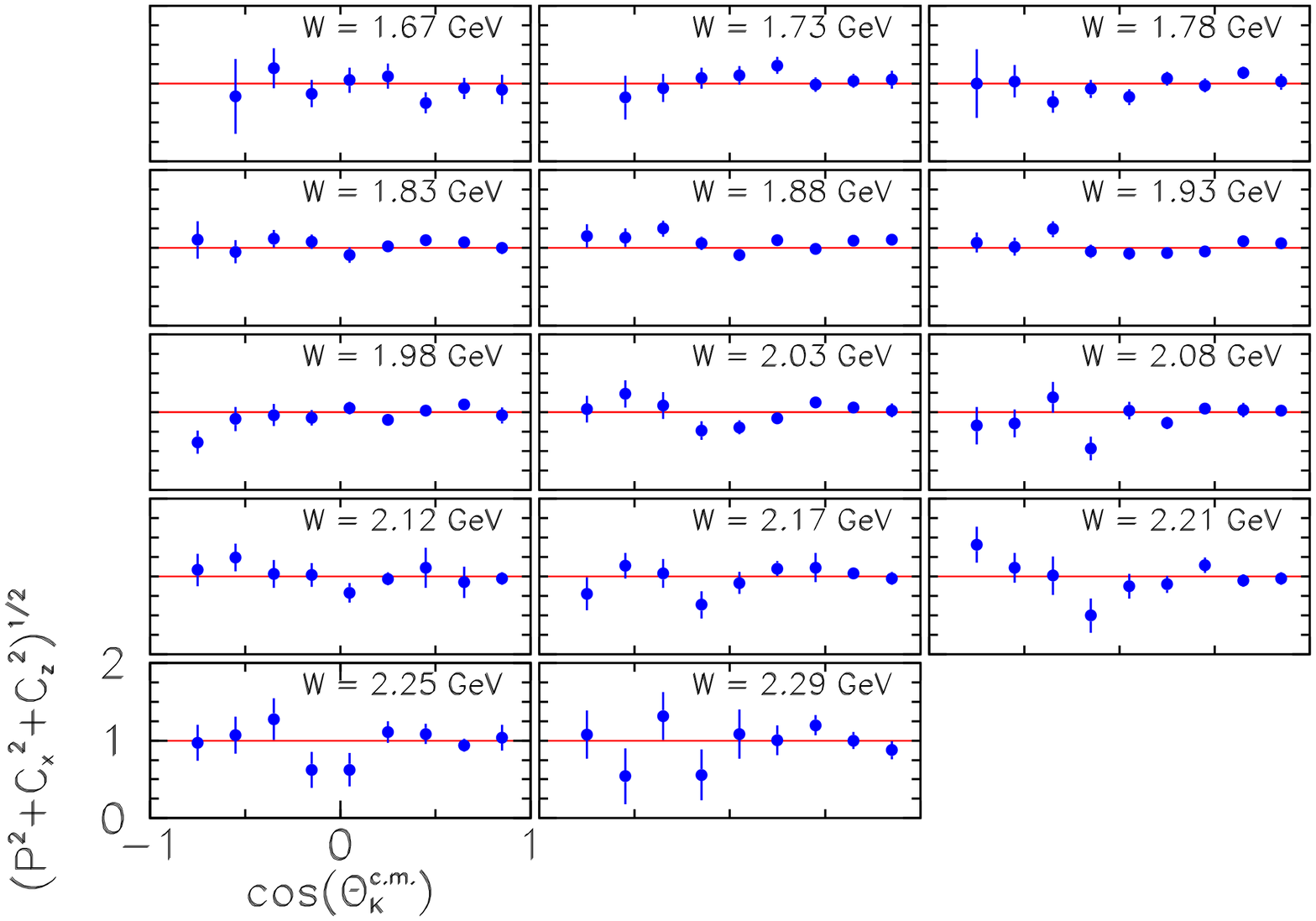}}
\caption{(Color online) 
The magnitude of the $\Lambda$ hyperon polarization observable vector 
$R_\Lambda=\sqrt{P^2 + C_x^2 + C_z^2}$ 
in the same binning as Figs.~\ref{fig:czlambda_angle} and
~\ref{fig:cxlambda_angle}.  $R_\Lambda$ is consistent with unity over all
values of $W$ and kaon angle.
}
\label{fig:rvalues}       
\end{figure*}

\begin{figure*}
\resizebox{1.2\textwidth}{!}{\includegraphics[10,50][700,500]{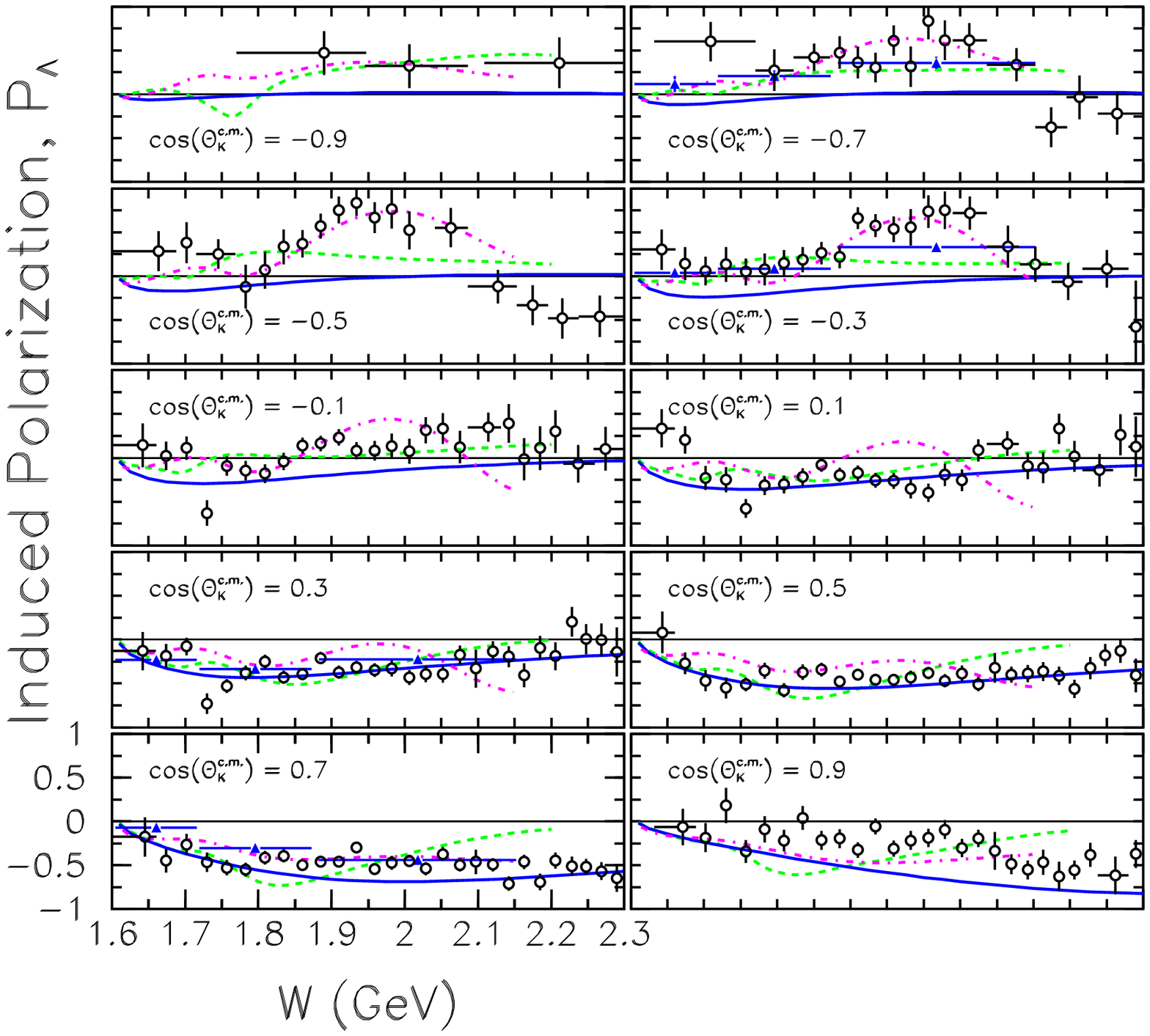}}
\caption{(Color online) Induced recoil polarization, $P$, of the
$\Lambda$ hyperon in $\gamma +p \to K^+ + \vec\Lambda$.  Open circles
(black) from Ref.~\cite{mcnabb}, triangles (blue) from
Ref.~\cite{bonn2}.  The dashed (green) curves are from
Kaon-MAID~\cite{maid}, 
the dot-dashed
(magenta) curves are from GENT~\cite{jan}, and
the solid (blue) curves are the Regge model GLV~\cite{lag1, lag2}.  }
\label{fig:pdata}       
\end{figure*}

Figure~\ref{fig:rvalues} displays the values for $R_\Lambda$ for the
$\Lambda$ hyperons obtained when combining the present results with
those of McNabb {\it et al.}~\cite{mcnabb}. The binning is the same as
for Figs.~\ref{fig:czlambda_angle} and~\ref{fig:cxlambda_angle}, with
the upper limit of $W=2.29$ GeV set by the range of the
previously-published data for $P$.  For ease of comparison we include
the previously-published data for $P$ in Fig.~\ref{fig:pdata}.  The
data in Fig.~\ref{fig:rvalues} combine the present $C_x$ and $C_z$
results with $P$ values interpolated to closely match the present $W$
and kaon angle bins.  The error bars are given by standard error
propagation, approximating the uncertainties on $C_x$ and $C_z$ as
statistically independent.

It is striking how close the magnitude of $R_\Lambda$ is to its
maximum possible value of +1 across all values of $W$ and kaon angle.
Taking the weighted mean over the data at all energies and angles we
find
\begin{equation}
\overline{R}_\Lambda = 1.01 \pm 0.01.
\end{equation}
This is consistent with unity within the given statistical uncertainty
on the mean, and certainly within our stated systematic uncertainty on
the beam polarization.  Some data points exceed the maximum allowed
value of unity by several sigma, but this must be expected on
statistical grounds.  The $\chi^2$ for a fit to the hypothesis that
$R_\Lambda = 1$ is 145 for 123 degrees of freedom, for a reduced
chi-square of 1.18, which is a good fit.  Thus, the deviations are
probably dominated by random measurement errors.

One may therefore conclude that the $\Lambda$ hyperons produced in
$\vec\gamma + p \to K^+ + \vec\Lambda$ with circularly polarized
photons appear 100\% spin polarized.  Since this situation is not {\it
required} by the kinematics of the reaction, there must be some as yet
unknown dynamical origin of this phenomenon.

The $\Lambda$ polarization direction is determined largely by the
photon helicity direction, since generally $C_z$ is the largest
component.  Careful examination of Figs.~\ref{fig:czlambda_w}
and~\ref{fig:pdata} shows where the induced polarization $P$ ``fills
in'' missing strength of $C_z$.  For example, at forward angles and
high energies, $C_z$ is reduced from unity, easily seen in the bottom
right panel of Fig.~\ref{fig:czlambda_w}, but the induced polarization
$P$ is large and negative in Fig.~\ref{fig:pdata}.  As another
example, near $W= 2.08$ GeV and $\cos\theta_{K^+}^{c.m.} = -0.55$, $P$
is large and positive just where $C_z$ dips down to about $+0.2$ and
$C_x$ is at $-0.6$.

\subsection{\label{sec:relation}Possible relation between $C_x$ and $C_z$}

Looking at the results shown in Figs.~\ref{fig:czlambda_w} 
through~\ref{fig:cxlambda_angle} suggests an empirical relation
between $C_x$ and $C_z$, specifically
\begin{equation}
C_z \simeq C_x + 1.
\label{eq:cxczone}
\end{equation}
Taking the weighted mean of $D \equiv C_z-C_x-1$ over all
values of $W$ and kaon angle leads to the value $\overline D = 0.054
\pm 0.012$.  In this case the $\chi^2$ for a fit to the hypothesis of
Eq.~\ref{eq:cxczone} is 306 for 159 degrees of freedom, or 1.92 for the
reduced $\chi^2$.  This is a poor fit, so our confidence in the
accuracy of this simple empirical relationship is limited, and
indicates that it needs experimental confirmation.  We can offer no
explanation for this curious relationship.  Linearity between $C_x$
and $C_z$ suggests rotating the coordinate axes by $+\pi/4$ about the
$\hat y$ axis, such that $C_x$ and $C_z$ are mapped onto two new axes,
$C_1$ and $C_2$.  The new variable $C_2$ would be approximately
constant with a value of $1/\sqrt{2}$ and all the variation with $W$
and kaon angle would be in $C_1$.  $C_2$ would represent a helicity
dependent but otherwise constant contribution to the cross section,
while $C_1$ would contain dynamical information.  In that case, the
three observables $C_x, C_z$, and $P$ would be reduced to a single
independent quantity.  One could define a phase angle, $\psi$, between
the induced and the transferred polarizations as $\psi=\tan^{-1} P/C_1
$.  The two relationships from Eqs.~\ref{eq:are} and~\ref{eq:cxczone},
together with $\psi$, would specify all three components of the
$\Lambda$ polarization.  The limited statistical precision of the
present results precludes drawing a stronger conclusion here.

\subsection{\label{sec:modelscomp}Comparison to Hadronic Models}

The results are compared in Figs.~\ref{fig:czlambda_w}
through~\ref{fig:cxsigma0_angle} with a group of recent calculations
based on published models.  It should be noted that none of these
calculations were refitted for the purpose of matching these new data.
In that sense, the curves shown in these figures are extrapolations of
the models to previously unmeasured observables.

First consider some recent effective-Lagrangian models of hyperon
photoproduction that evaluate tree-level Feynman diagrams including
resonant and non-resonant exchanges of baryons and mesons.  The
advantages of the tree-level approach, {\it i.e.} to not include the
effects of channel coupling and rescattering, are to limit complexity
and to identify the dominant trends.

For $K^+\Lambda$ production, the model of Mart and
Bennhold~\cite{mart} has four baryon resonance contributions.  Near
threshold, the steep rise of the cross section is accounted for with
the $N^*$ states $S_{11}(1650)$, $P_{11}(1710)$, and
$P_{13}(1720)$.  To explain the broad cross-section bump in the mass
range above these resonances~\cite{bradforddsdo,bonn2}, they
introduced the $D_{13}(1895)$ resonance that was predicted in the
relativized quark models of Capstick and Roberts~\cite{cap} and
L\"oring, Metsch, and Petry~\cite{loring} to have especially strong
coupling to the $K^+\Lambda$ channel.  In addition, the higher mass
region has contributions, in this model, from the exchange of vector
$K^*(892)$ and pseudovector $K_1(1270)$ mesons.  The hadronic form
factors, cutoff masses, and the prescription for enforcing gauge
invariance were elements of the model for which specific choices were
made.  The content of this model is embedded in the Kaon-MAID
code~\cite{maid} which was used for the comparisons in this paper.
This model was fitted to preliminary results from the experiment at
Bonn/{\small SAPHIR}~\cite{bonn1}, and offers a fair description of
those results.

Analysis by Saghai {\it et al.}~\cite{saghai1} using the same cross
section data showed that, by tuning the background processes involved,
the need for the extra resonance was removed. Also, Janssen {\it et
al.}~\cite{jan,jan_a} (designated GENT here) showed that the same data
set was not complete enough to make firm statements since models with
and without the presence of the hypothesized $N^*(1895) D_{13}$ resulted
in equally good fits to the data. A subsequent related analysis~\cite{ireland}
which also fitted to photon beam asymmetry measurements
from SPring-8~\cite{zegers} and electroproduction data measured at
Jefferson Lab~\cite{mohring}, indicated weak evidence for one or more
of $S_{11}$, $P_{11}$, $P_{13}$, or $D_{13} (1895)$, with the $P_{11}$
solution giving the best fit. The conclusion was that a more
comprehensive data set would be required to make further progress.

Recently, more elaborate model calculations have been undertaken that
consider amplitude-level channel coupling or at least simultaneous
fitting to several incoherent reaction channels. Penner and
Mosel~\cite{penner} found fair agreement for the $K^+\Lambda$ data
without invoking a new $D_{13}$ structure.  Chiang {\it et
al.}~\cite{chiang} showed that coupled channel effects are significant
at the $20\%$ level in the total cross sections when including pionic
final states.  Shklyar, Lenske, and Mosel~\cite{shklyar} (designated
SLM here) used a unitary coupled-channel effective Lagrangian model
applied to $\pi$ and $\gamma$ -induced reactions to find dominant
resonant contributions from $S_{11}(1650)$, $P_{13}(1720)$, and
$P_{13}(1895)$ states, but not from $P_{11}(1710)$ or
$D_{13}(1895)$. Their conclusion held despite the discrepancies
between previous cross section data from CLAS~\cite{mcnabb} and
{\small SAPHIR}~\cite{bonn2}.

A dynamical coupled-channel model of $K^+\Lambda$ photoproduction
which emphasized intermediate $\pi N$ states was presented by
Julia-Diaz {\it et al.}~\cite{saghai2} (designated SAP here).  The
model was constrained by results for the hadronic $\pi N \to K Y$
channels.  To avoid duality issues, $t$-channel exchange was limited
only to non-resonant $K$ exchange.  Using published
photoproduction~\cite{bradforddsdo,bonn2} and hadronic cross section
data, and the $\Lambda$ polarization data~\cite{mcnabb, zegers,
althoff}, they sought the dominant baryon resonance contributions to
$K^+\Lambda$ photoproduction.  The model demonstrated dominant
contributions from the $N^*$ states $S_{11}(1535)$, $P_{13}(1900)$, and
$D_{13}(1520)$.  Contributions from three new nucleon resonances were
found to be significant, specifically $D_{13}(1954)$, $S_{11}(1806)$,
and $P_{13}(1893)$.  The model showed significant sensitivity to
induced polarization $P$ of the $\Lambda$, so one may expect similar
sensitivities in $C_x$ and $C_z$.

A partial wave analysis of the combined data sets for the reactions
$\gamma p \to \pi N, \eta N, K^+\Lambda, K^+\Sigma^0, K^0\Sigma^+$ has
been reported by a group from Bonn, Gatchina, and
Giessen~\cite{sarantsev, anisovich, anisovich2} (designated BG here).
The method used a relativistically invariant operator expansion method
with relativistic Breit-Wigner representations of selected resonances
and reggeized $t$-channel exchanges.  Some close-in-mass resonances
were coupled using a K-matrix formalism, but overall unitarity
violation was allowed.  The analysis included the differential cross
sections, beam asymmetry for the $\eta$ and the $\Lambda$ cases, and
induced recoil polarizations $P$ for the $\Lambda$ and the $\Sigma^0$.
We note that the $KY$ CLAS cross section data used in the fits were
from Ref.~\cite{mcnabb}, and not the newer and more complete results
from Ref.~\cite{bradforddsdo}.  Compared to other recent models, BG
takes into account a larger range of experimental information
simultaneously.  The spin observables were found to be vital to
extract the signatures of resonances as revealed by their mutual
interferences.  Strong evidence was found for several new $N^*$ states
including $P_{11}(1840)$ and $D_{13}(1875)$, with weaker evidence for
a $D_{13}(2170)$.  It might be expected that ``new'' resonances that
coupled significantly to $KY$ and are seen via their effect on spin
observables should also have a significant impact on $C_x$ and $C_z$.

In another recent approach, Corthals, Ryckebusch, and Van
Cauteren~\cite{corthals} used a ``Regge plus resonance'' (RPR) picture
to reproduce the CLAS differential cross sections~\cite{bradforddsdo},
recoil polarizations~\cite{mcnabb}, and LEPS beam
asymmetries~\cite{zegers} for $K^+\Lambda$ production.  By fixing the
few parameters of a Regge model of $K^*$ and $K$ exchange at energies
between 5 and 16 GeV, they found four acceptable ways of describing
the available high energy data~\cite{slac}.  They evolved these
solutions into the nucleon resonance region as a way to describe the
``background'' to the $K^+\Lambda$ baryon resonance production cross
section.  Despite concerns about breaking duality, the advantage of
this approach is the relatively small number of free parameters that
are needed when compared to $s$-channel dominated isobar models.  The
latter generally require evaluation of many more diagrams, even at
tree level, to approach the measured cross sections.  A standard group
of ``core'' resonances was included, the $S_{11}(1650)$, the
$P_{13}(1720)$, and the $P_{11}(1710)$, together with a small set of
extra $N^*$ resonances.  Three acceptable fits to the data were
obtained. The set of additional $N^*$ resonances tested were a
$P_{13}(1900)$, a $P_{11}(1900)$, and a $D_{13}(1900)$.  Remarkably,
one satisfactory solution required no additional baryon resonances at
all.  The other solutions showed the need for a $P(1900)$ resonance,
but the $D_{13}(1900)$ hypothesis did not lead to better fits.  The
authors concluded that the experimental information is still not
precise enough to make an unambiguous case for the resonance
contribution(s) in the 1900 MeV mass range.  However, a shortcoming of
this RPR approach is that it only works for the forward angle region
where the Regge parameterization of the cross section can be expected
to work.  Much of the sensitivity to resonance contributions that
shows up more strongly at mid and back angles is thus ignored.  It is
of interest, therefore, to see how the extrapolations of these RPR
solutions, with no additional fitting, match the observables reported
in this paper.

Although it is to be expected that $s$-channel resonance structure is a
significant component of the $K^+\Lambda$ and $K^+\Sigma^0$ reaction
mechanisms, it is instructive to compare to a model that has no such
content at all.  The model of Guidal, Laget, and
Vanderhaeghen~\cite{lag1,lag2} (GLV) is such a model, in which the exchanges
are restricted to two linear Regge trajectories corresponding to the
vector $K^*$ and the pseudovector $K_1$.  The model was fit to
higher-energy photoproduction data where there is little doubt of the
dominance of these exchanges.  In this paper, we extend that model
into the resonance region in order to make a critical comparison.

Having introduced the recent models of hyperon photoproduction, we
proceed with some remarks on their behavior in relation to the present
results.  The models have in common that at threshold the values are
$C_z = +1.0$ and $C_x = 0.0$, which is as expected on the basis of the
naive picture introduced above in which there is no orbital angular
momentum available to carry off any of the $\hat z$ component of
angular momentum.  The exception is the Kaon-MAID model~\cite{maid}
which clearly contains a sign error, since it starts at $C_z = -1.0$
at threshold.  We chose not to reverse this sign by hand but to show
the model curve exactly as it is publicly available.  Furthermore,
Fig.~\ref{fig:czlambda_angle} shows that the BG, SAP, and SLM models
correctly show that $C_z \to +1.0$ at the extreme scattering angles
$\cos \theta_{K^+}^{c.m.} \to +1.0$.  This must be the case since the
$z$ component of angular momentum must be conserved via the hyperon
spin in this limit.  In the same angle limit $C_x\to 0$ and all models
exhibit this correctly.  For $\cos \theta_{K^+}^{c.m.} \to -1.0$ the
same limits hold again, and the RPR, BG, and SAP models show this
correctly, while GENT appears not to extrapolate to these limits.

The next remark is that none of the existing models can be said to do
even a fair job predicting the behavior of $C_z$ and $C_x$ anywhere
away from threshold.  Only the older model GENT of Janssen {\it et
al.}~\cite{jan,jan_a} approximates the qualitative finding that $C_z$
is large and positive over most of the measured range.  The follow-on
model of RPR~\cite{corthals} is less successful by comparison.  It is
notable that the pure Regge GLV model~\cite{lag1,lag2}, containing
only two trajectories and no parameters adjusted to fit the
resonance-region data, does no worse than the much more elaborate
hadrodynamic models.

We take the poor agreement of existing reaction models with the
results as an indication that all models will be able to use these
results to refine their contents.

\subsection{\label{sec:pqcd}Comparison to pQCD Limits}

Afanasev, Carlson, and Wahlquist~\cite{afanasev} studied polarized
parton distributions via meson photoproduction in a model where pQCD
was used to describe direct photoproduction of a meson from a quark.  The
approach is applicable for high transverse momenta where short-range
processes are dominant.  It was used in the analysis of the
reaction $p(\vec \gamma,\vec p)\pi^0$ with circularly polarized
photons
in Ref.~\cite{wijesooriya}.
Assuming helicity conservation, this model predicted
\begin{equation}
P = C_{x^\prime} = 0
\label{pcx0}
\end{equation}
and
\begin{equation}
C_{z^\prime} = \frac{s^2 - u^2}{s^2 + u^2}
\label{eqn:czsu}
\end{equation}
in the $\{x^\prime,z^\prime\}$ basis of Fig.~\ref{fig:axes}, where
$s$, $t$, and $u$ are the usual Mandelstam variables.  In the limit of
massless quarks $C_{z^\prime} \to 0$ as $|t|\to 0$, and $C_{z^\prime}
\to 1$ when $|u|\to 0$ at large angles and large $|t|$.  The model
further assumes the polarization of the struck quark is the same as
the polarization of the outgoing hyperon, undiluted by hadronization
effects.  In the present discussion of $p(\vec\gamma,\vec\Lambda) K^+$,
the strange quark is expected to carry the $\Lambda$ spin as expected in
the quark model.  The ``short-range process'' involves the creation of
an $s \overline s$ quark pair.  The light-cone momentum fraction of
the active quark, $x$, is defined~\cite{afanasev} for photoproduction
as
\begin{equation}
x = \frac{-t}{s+u}.
\end{equation} 
In the present measurements we have $0.06 < x < 0.6$.  Thus, we span
the regime where the struck quark could be a strange sea quark, which
hadronizes into a $\Lambda$ hyperon while the anti-strange quark
produces the kaon.  But at large $|t|$ where this approach could be
valid we are in the valence quark regime.

\begin{figure}
\resizebox{0.50\textwidth}{!}{\includegraphics{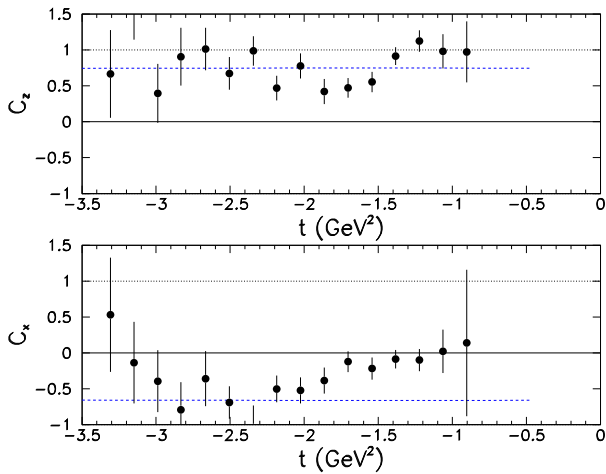}}
\vspace{-1.0cm}
\caption{(Color online) The observables $C_z$ and $C_x$ for the
reaction $\vec\gamma + p \to K^+ + \vec\Lambda$, plotted as a function
of $t$.  The dashed (blue) curves are a prediction~\cite{afanasev,
wijesooriya} \ from perturbative QCD assuming helicity conservation at
the quark level.  }
\label{fig:pqcd}       
\end{figure}

Since our results show that $C_z$ is large and positive over most of
our kinematic range, it is clear that quark helicity in the baryon is not
conserved in this reaction.  Nevertheless one can look at the
kinematic range where Eq.~\ref{eqn:czsu} is thought to be most
applicable.  Figure~\ref{fig:pqcd} shows our results for the largest
$|t|$ values measured, stemming from $\cos\theta_{K^+}^{c.m.} =
-0.75$, as a function of $t$.  In the limit of large kaon angle,
helicity conservation requires $C_{z^\prime}$ to approach unity with our
axis definition.  Rotating the prediction to yield $C_x$ and $C_z$
results in the dashed lines in the figure.  The agreement with the
model is fair to good at large values of $|t|$.  Whether or not this
is fortuitous is uncertain, since the domain of applicability of the
model is not well defined and non-perturbative effects clearly
dominate the data at lower $|t|$.

Thus, the correct interpretation of this reasonable agreement with the
model is not clear.  The partial success of this model for the present
results on $K^+ \vec \Lambda$ production is in contrast to its
complete failure when applied to $\pi^0 \vec p$ photoproduction
~\cite{wijesooriya} in a similar range of $W$.  In that measurement
the recoiling protons are always much less polarized than the pQCD
model suggested.

\subsection{\label{sec:electro}Comparison to Electroproduction}

The present results for photoproduction can be compared to previous
measurements for the reaction $p(e,e^\prime K^+\Lambda)$ made by
CLAS~\cite{carman}.  Additional observables arise in electroproduction
on account of the extra spin degrees of freedom associated with the
virtual photons at finite values of $Q^2$.  However, the formalism of
the electroproduction structure functions merges smoothly into the
limiting case of photoproduction at $Q^2 = 0$ (GeV/c)$^2$, as written
explicitly, for example, in Ref.~\cite{knoechlein}.  The
electroproduction results were averaged over the range $0.3 < Q^2 <
1.5$ (GeV/c)$^2$ and also averaged over the azimuthal angle between
the electron scattering and the hadronic reaction planes.  The
transferred polarization component along the direction of the virtual
photon, called $P^\prime_z$ in Ref.~\cite{carman}, is large (between
+0.6 and +1.0) and roughly independent of the kaon angle for values of
$W$ at 1.69, 1.84, and 2.03 GeV.  There is a mild trend toward smaller
values of $P^\prime_z$ with increasing kaon angle.  This is consistent
with our findings discussed above, where $C_z$ is close to +1.0 for
the same $W$ values and across all kaon angles, as seen in
Fig.~\ref{fig:czlambda_angle}.  In the electroproduction measurement
the orthogonal $\hat x$ axis was chosen in the electron scattering
plane, while in the present paper we can only choose it in the
hadronic reaction plane. However, we note that the corresponding
$P^\prime_x$ values in electroproduction are small ($ < +0.2$) across
all kaon angles and $W$ values.  This is again in qualitative
agreement with our observed values of $C_x$.  Thus, we can conclude
that the photo- and electro- production measurements show the same
qualitative behavior, meaning that there is no rapid departure from
the photoproduction systematics as one moves out in $Q^2$ from zero to
about 1.5 (GeV/c)$^2$.

\subsection{\label{sec:sigmaresults}Results for the $\Sigma^0$}

\begin{figure*}
\vspace{1.0cm}
\resizebox{0.8\textwidth}{!}{\includegraphics[70,50][600,450]{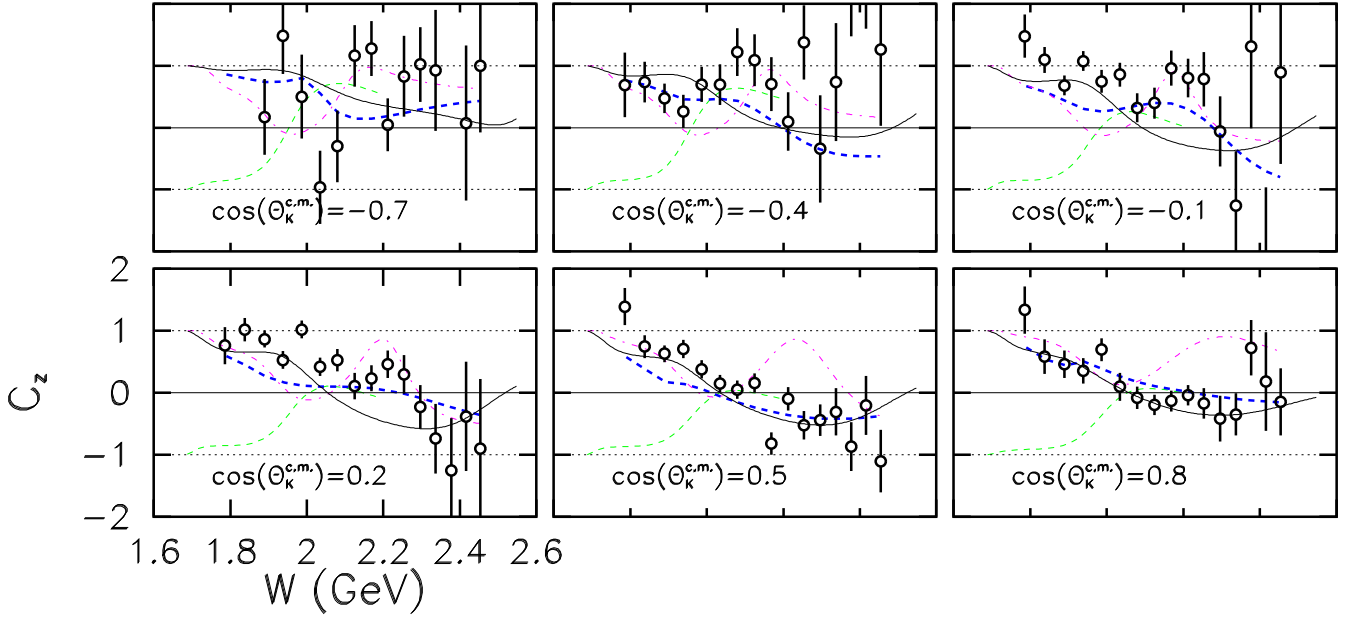}}
\vspace{-4.0cm}
\caption{(Color online) The observable $C_z$ for the reaction 
$\vec\gamma + p \to K^+ + \vec\Sigma^0$, 
plotted as a function of the c.m. energy $W$.  
The circles are the results of this measurement, 
with uncertainties discussed in the text.
The thin dashed (green) curves are from Kaon-MAID~\cite{maid}, 
the thick dashed (blue) curves are from BG~\cite{sarantsev2},  
the thin solid (black) curves are from RPR~\cite{corthals}, and
the thick dot-dashed (magenta) curves are from GENT~\cite{jan}.  
}
\label{fig:czsigma0_w}       
\end{figure*}

\begin{figure*}
\vspace{1.0cm}
\resizebox{0.8\textwidth}{!}{\includegraphics[70,50][600,450]{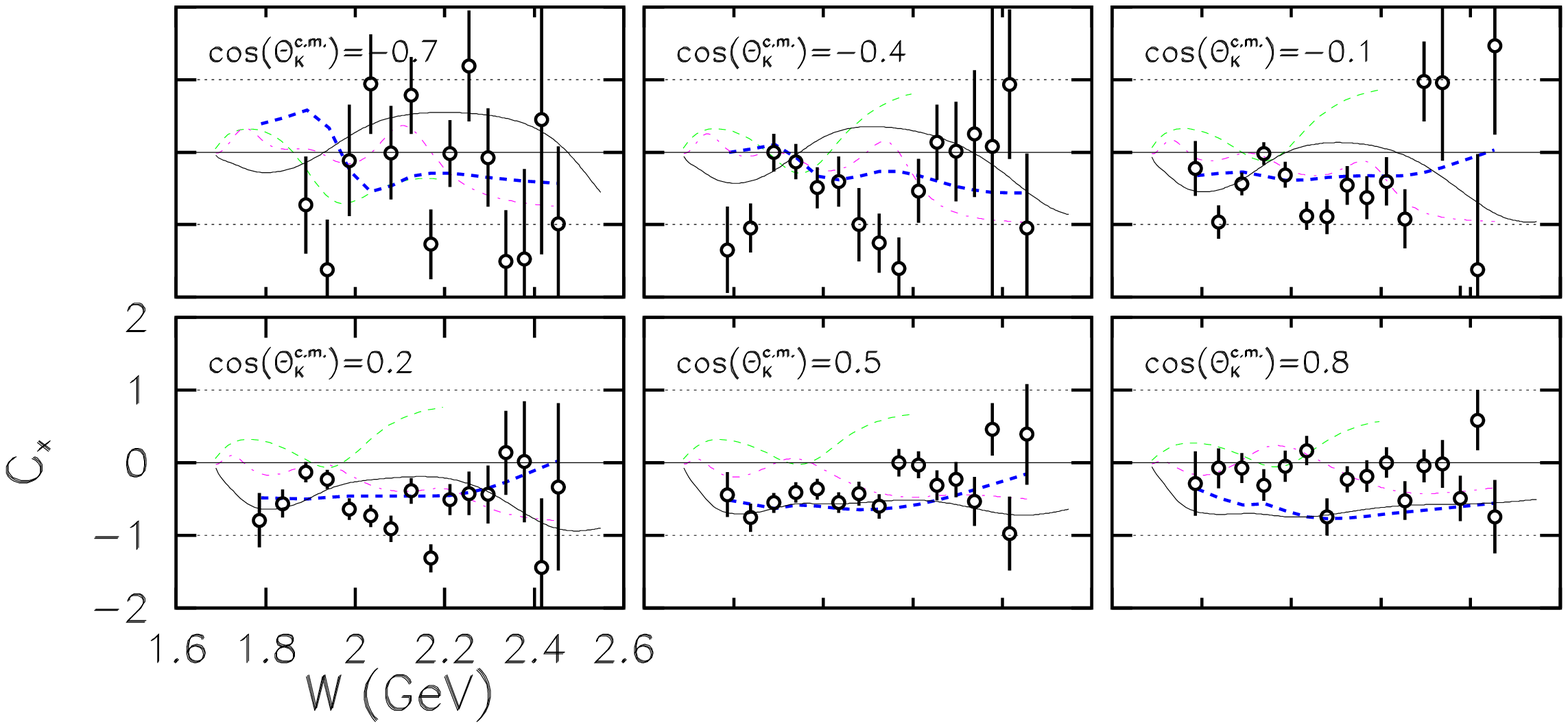}}
\vspace{-4.0cm}
\caption{(Color online) The observable $C_x$ for the reaction
$\vec\gamma + p \to K^+ + \vec\Sigma^0$, 
plotted as a function of the
c.m. energy $W$.  
The circles are the results of this measurement, 
with uncertainties discussed in the text.  
The thin dashed (green) curves are from Kaon-MAID~\cite{maid}, 
the thick dashed (blue) curves are from BG~\cite{sarantsev2},  
the thin solid (black) curves are from RPR~\cite{corthals}, and
the thick dot-dashed (magenta) curves are from GENT~\cite{jan}.  
}
\label{fig:cxsigma0_w}       
\end{figure*}

\begin{figure*}
\resizebox{0.6\textwidth}{!}{\includegraphics[70,50][600,500]{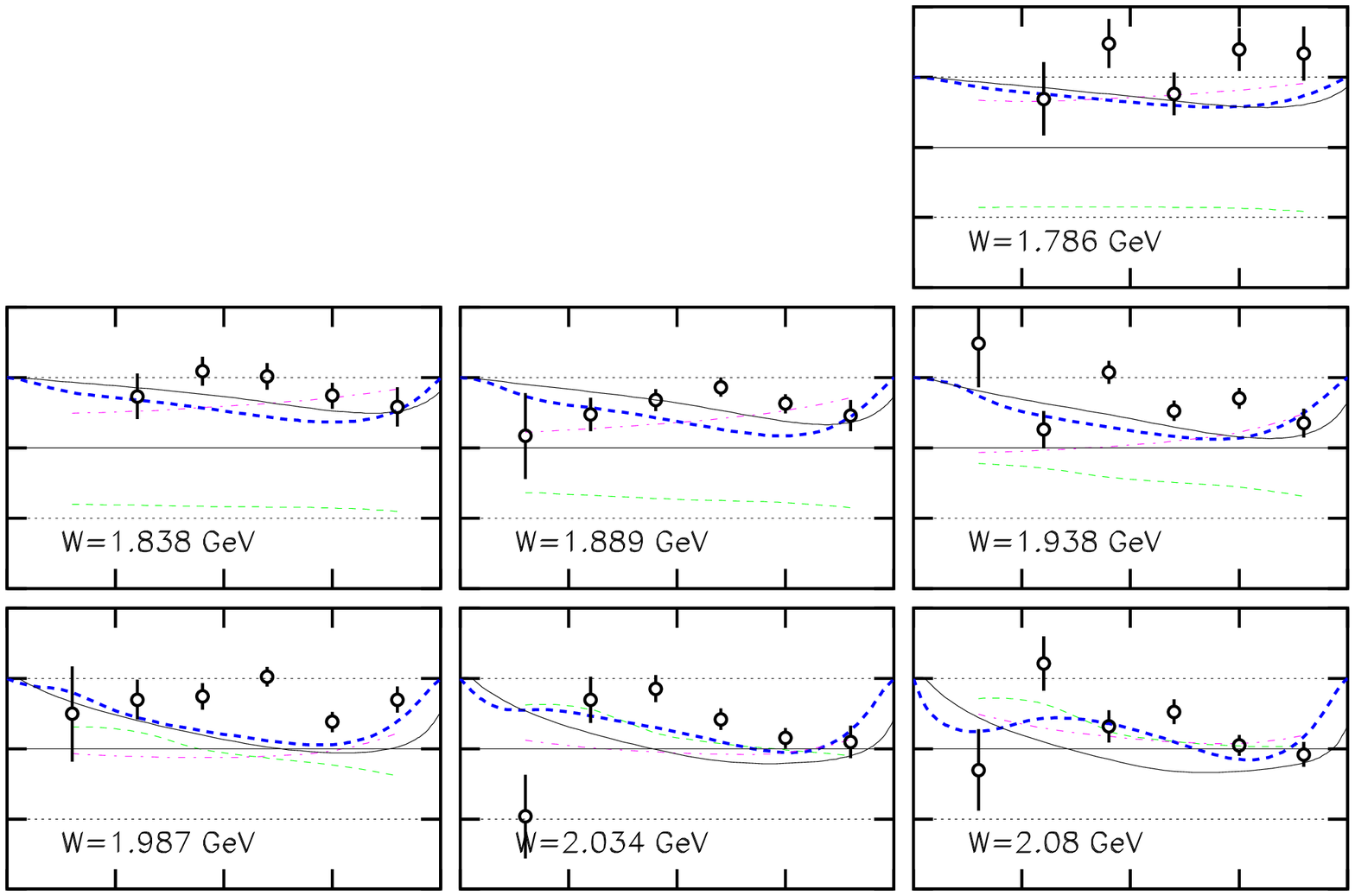}}
\resizebox{0.6\textwidth}{!}{\includegraphics[70,50][600,440]{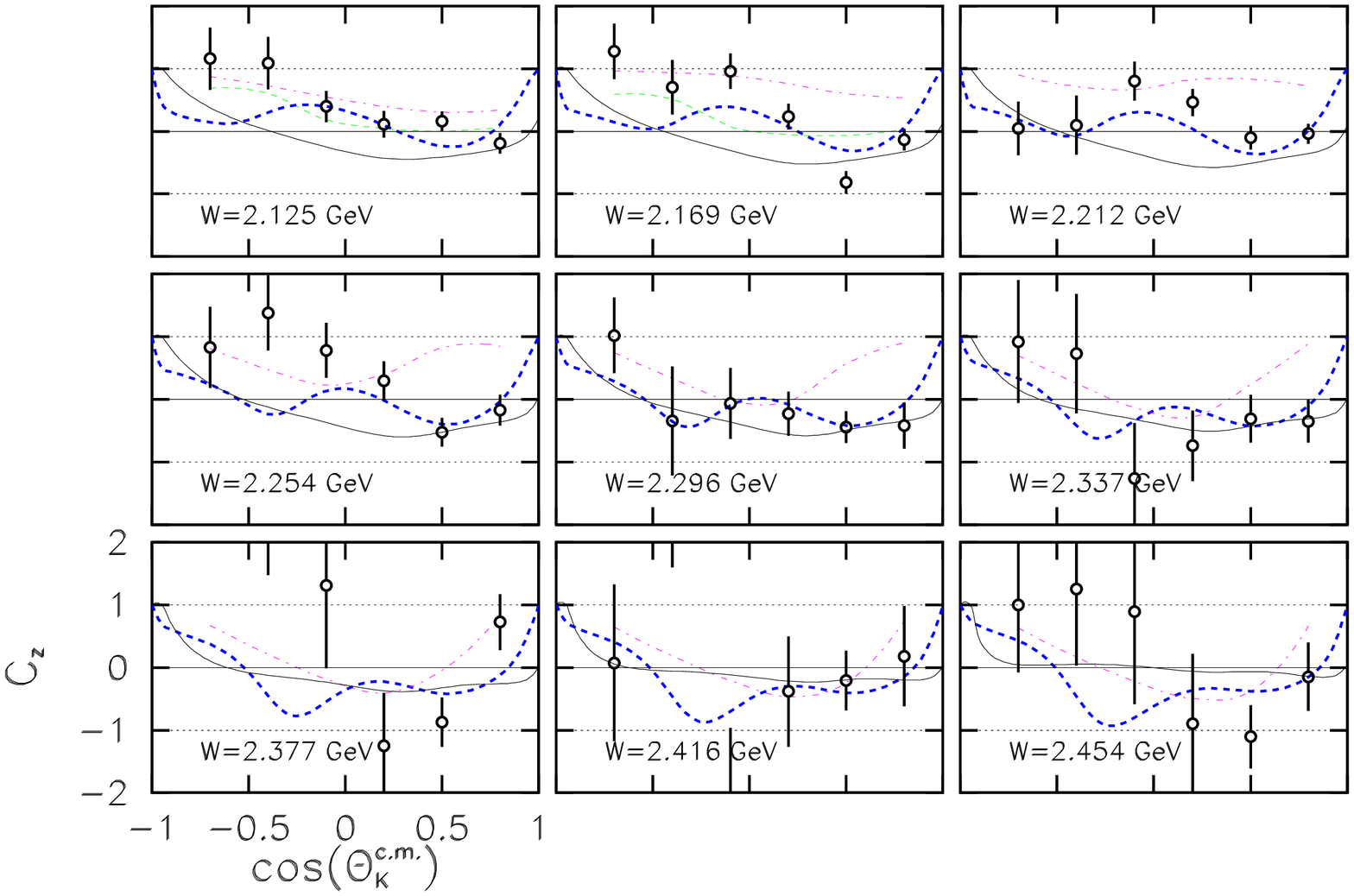}}
\caption{
(Color online) The observable $C_z$ for the reaction $\vec\gamma + p
\to K^+ + \vec\Sigma^0$, plotted as a function of the kaon angle.
The 16 panels are for increasing values of $W$ in steps of about 50 MeV.  
The circles are the results of this measurement, with uncertainties discussed in the text.  
The thin dashed (green) curves are from Kaon-MAID~\cite{maid}, 
the thick dashed (blue) curves are from BG~\cite{sarantsev2},  
the thin solid (black) curves are from RPR~\cite{corthals}, and
the thick dot-dashed (magenta) curves are from GENT~\cite{jan}.  
}
\label{fig:czsigma0_angle}       
\end{figure*}

\begin{figure*}
\resizebox{0.6\textwidth}{!}{\includegraphics[70,50][600,500]{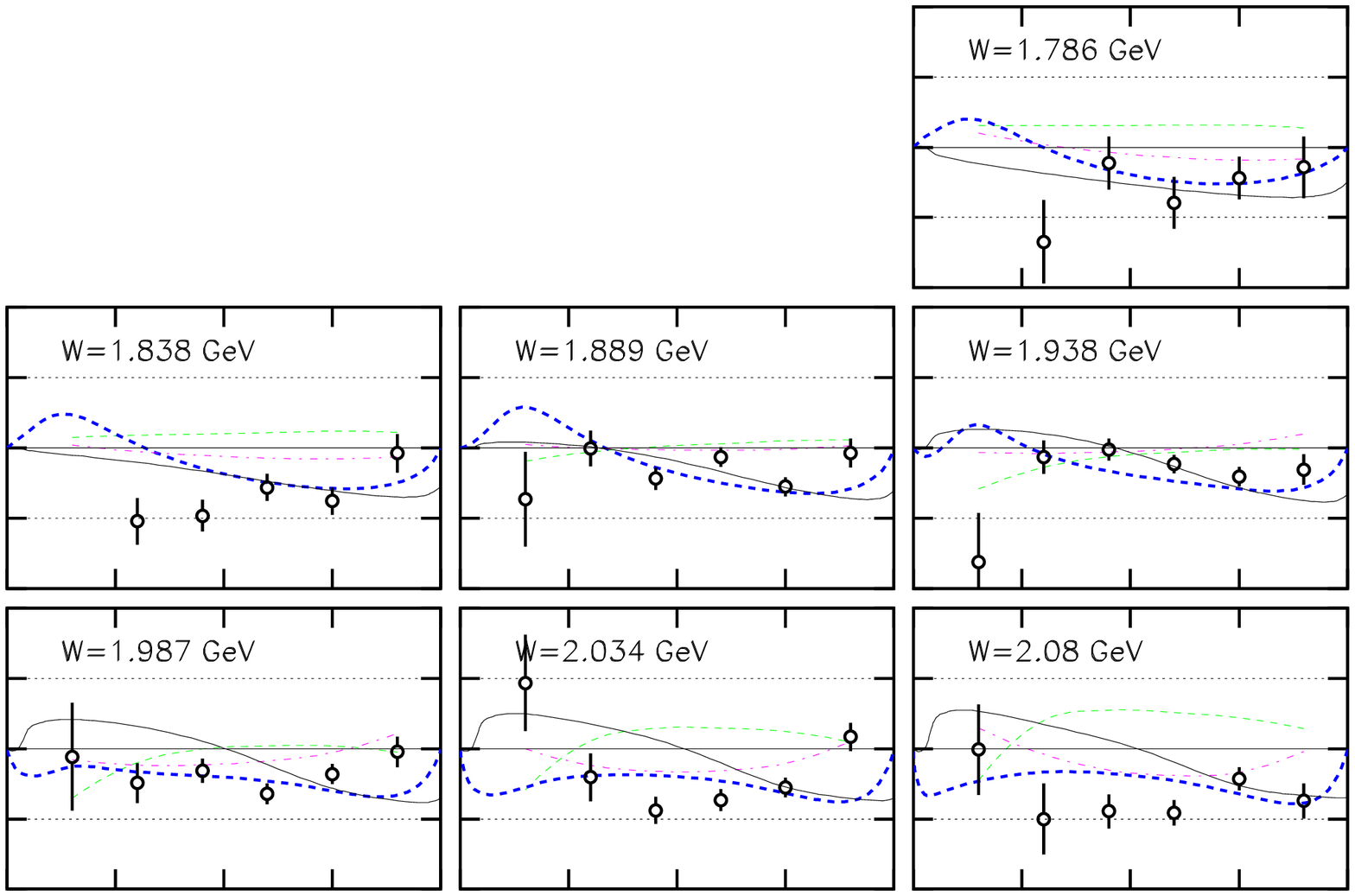}}
\resizebox{0.6\textwidth}{!}{\includegraphics[70,50][600,440]{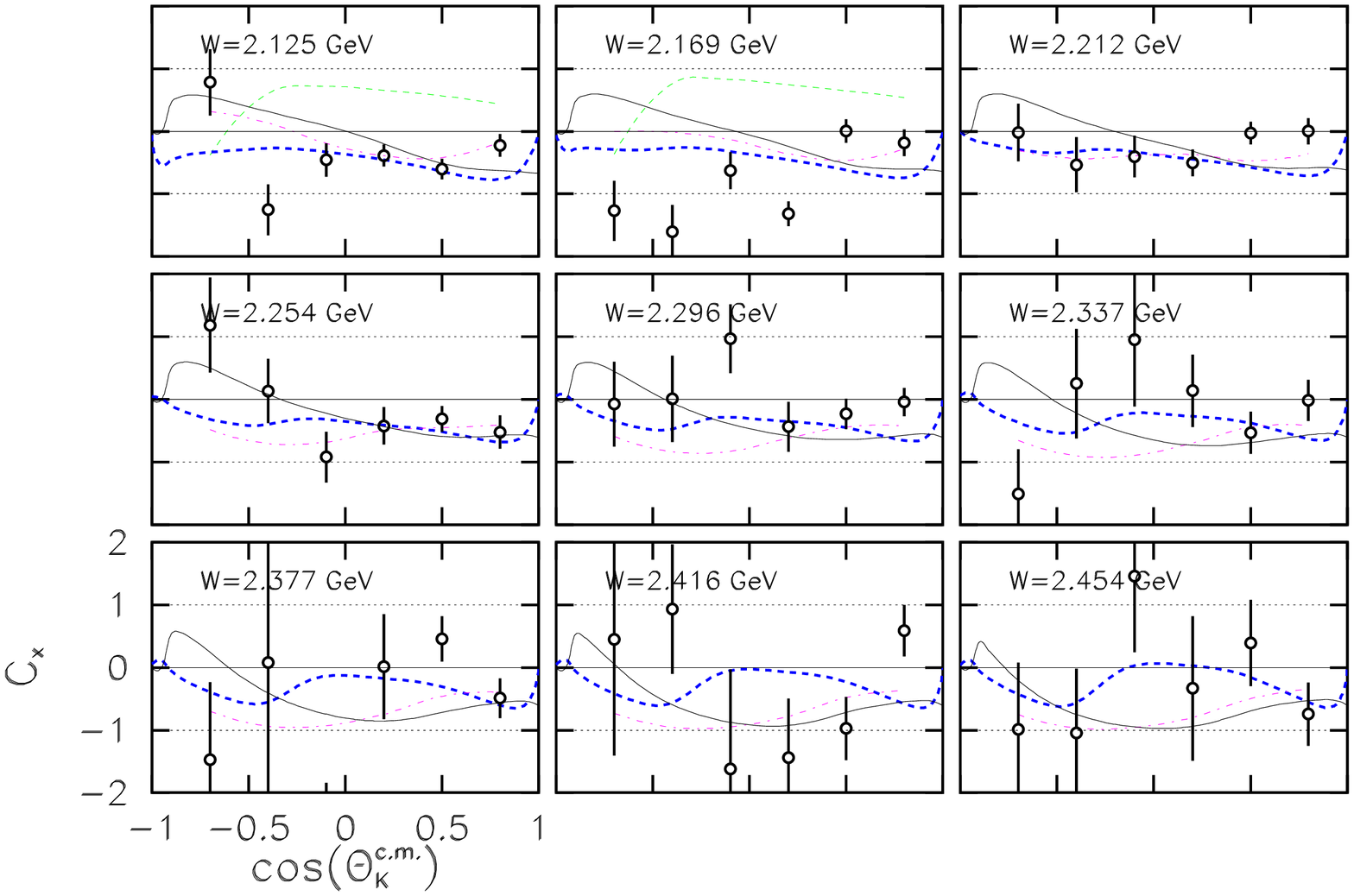}}
\caption{
(Color online) The observable $C_x$ for the reaction $\vec\gamma + p
\to K^+ + \vec\Sigma^0$, plotted as a function of the kaon angle.
The 16 panels are for increasing values of $W$ in steps of about 50 MeV.  
The circles are the results of this measurement, with uncertainties discussed in the text.  
The thin dashed (green) curves are from Kaon-MAID~\cite{maid}, 
the thick dashed (blue) curves are from BG~\cite{sarantsev2},  
the thin solid (black) curves are from RPR~\cite{corthals}, and
the thick dot-dashed (magenta) curves are from GENT~\cite{jan}.  
}
\label{fig:cxsigma0_angle}       
\end{figure*}

In the quark model the {\it ud} quarks in the $\Sigma^0$ are in a spin
triplet state instead of a spin singlet as in the $\Lambda$.  The
created strange quark is not alone in determining the spin of the
overall hyperon in the $\Sigma^0$.  Thus one may expect the behavior
of $C_x$ and $C_z$ for the $\Sigma^0$ to differ from that of the
$\Lambda$.  Figures~\ref{fig:czsigma0_w} to \ref{fig:cxsigma0_angle}
present these results, and indeed it is immediately clear that the
trends in this case are not the same as in the previous discussion.
Note first that only 6 kaon angle bins were used, centered at
$\cos \theta_{K^+}^{c.m.} = -0.7$ to $+0.8$ in steps of $0.3$.  This
was necessitated by the reduced sensitivity to the $\Sigma^0$
polarization due to the previously-discussed dilution caused by the
$\Sigma^0\to\gamma\Lambda$ decay.  Despite coarser binning, the
statistical precision of the $\Sigma^0$ results is still less good
than the $\Lambda$ results by a factor of 2 to 3.

The most dramatic differences can be seen comparing the
forward-hemisphere values of $C_z$ for the $\Sigma^0$ in
Fig.~\ref{fig:czsigma0_w} with the $\Lambda$ in
Fig.~\ref{fig:czlambda_w}.  Near $\cos \theta_{K^+}^{c.m.} = +0.45$,
$C_z$ for the $\Lambda$ is at unity for the whole range in $W$, while
for the $\Sigma^0$ it falls from $+1.0$ at threshold to large negative
values at the highest $W$.  The trends of the $C_x$ values for the
$\Sigma^0$ in Fig.~\ref{fig:cxsigma0_w} are, with limited statistical
precision, similar to those of the $\Lambda$ shown in
Fig.~\ref{fig:cxlambda_w}: $C_x$ is predominantly negative.  The
angular distributions for the $\Sigma^0$ in
Fig.~\ref{fig:czsigma0_angle} are compared to those for the $\Lambda$
in Fig.~\ref{fig:czlambda_angle}: the panels are placed to have the
same $W$ bins in the same location.  At $W=1.889$ GeV, for example,
the $\Sigma^0$ has a $C_z$ of about $+0.5$, while for the $\Lambda$ it
is at $+1.0$.  At $W=2.296$ GeV the $C_z$ for the $\Sigma^0$ is about
zero, while for the $\Lambda$ it is large and positive.  The
corresponding values of $C_x$ are similar between the two hyperons, as
seen in comparing Figs.~\ref{fig:cxsigma0_angle}
and~\ref{fig:cxlambda_angle}.



As was the case for the $\Lambda$ polarization, one expects that the
magnitude of the polarization transfer coefficients, $R_{\Sigma^0}$,
to be less than unity as per Eq.~\ref{eq:are}.  The lesser
statistical precision in the case of the $\Sigma^0$ for all three
components of the combination $\{C_x,P,C_z\}$ makes it more difficult
to compute this precisely.  However, we found that the angle and
energy averaged value is
\begin{equation}
\overline R_{\Sigma^0} = 0.82 \pm .03,
\end{equation}
which is clearly incompatible with the maximum possible value of
unity.  Thus, the $\Sigma^0$ cannot be said to be produced with
$100\%$ polarization from a fully polarized beam.  Thus, even if the
quark-level dynamics leading to the creation of an $s \overline s$
quark pair were the same in both the $\Lambda$ and $\Sigma^0$ reaction
channels, then the hadronization into a $\Lambda$ or a $\Sigma^0$
produces different final polarization states.  If the quark-level
dynamics are not relevant, one is left with the question of why the
$\Lambda$ is formed fully spin polarized but not so the $\Sigma^0$.

The previous remarks about the comparison to existing reaction models
apply to the $\Sigma^0$ case as well as the $\Lambda$ case.  While
none of the calculations can be said to agree well with the data, the
calculation of Corthals {\it et al.}~\cite{corthals} at least
reproduces the trend with $W$ at most angles, as shown in
Fig.~\ref{fig:czsigma0_w}.

\subsection{\label{sec:amplitudes}Further Discussion}

In addition to comparison to dynamical models, as done above, one can
ask what model-independent information is gained from these
measurements.  Photoproduction of pseudoscalar mesons from spin 1/2
baryons is described by four complex amplitudes that are functions of
the reaction kinematics~\cite{barker, goldstein, tabakin1}.  For
example, in the helicity basis where the photon has helicity $\pm1$,
one can easily enumerate four combinations of spins with overall
helicity flips of zero ($N$), one ($S_1$), one ($S_2$), or two ($D$)
units.  The letter notation is that of Barker {\it et
al.}~\cite{barker}.  In a transversity basis in which the proton and
hyperon have well-defined spin projections with respect to the $\hat
y$ axis normal to the reaction plane there are linear combinations of
the helicity amplitudes which are more convenient for studying
polarization observables~\cite{barker,adelseck,tabakin1}; they are
labeled $b_1, b_2, b_3, b_4$.  As shown in Table~\ref{table:amps},
these have an advantage that measurement of the cross sections
(designated $A$) plus the three single spin observables $\Sigma$,
$T$, $P$, yields the magnitudes of these four amplitudes.  The double spin
observables serve to define the three phases among the amplitudes.  We
note in passing that four CGLN amplitudes~\cite{cgln} form yet another
set of amplitudes that could be used~\cite{knoechlein}.
Table~\ref{table:amps} shows the algebraic relations among the
helicity and transversity amplitudes for the observables in hyperon
photoproduction presented in this paper.  At each value of Mandelstam
$s$ and $t$ there are seven real numbers and an arbitrary overall
phase which specify the scattering matrix.  All observable quantities
are expressible as bilinear products of the amplitudes, and thus there
are 16 observables.

\begin{table*}[ht]
\caption{
Amplitude combinations leading to the measured observables
in the helicity and transversity representations, adapted from
Ref.~\cite{barker}.  The axis convention is taken from that
reference, and is rotated from the one in this paper, as
discussed in the text.}
\vspace{0.5cm}
\centering
\begin{tabular}{|c|c|c|}
\hline
Observable                  & Helicity                      & Transversity     \\
                            & Representation                & Representation   \\
\hline\hline
$A$, $d\sigma/dt$           & $|N|^2+|S_1|^2+|S_2|^2+|D|^2$ &$|b_1|^2+|b_2|^2+|b_3|^2+|b_4|^2$  \\

$P$ $d\sigma/dt$            & $2 Im(S_2N^* - S_1 D^*)$      &$|b_1|^2-|b_2|^2+|b_3|^2-|b_4|^2$  \\

$C_{x^\prime}$ $d\sigma/dt$ & $-2 Re(S_2N^* + S_1 D^*)$     &$2 Im(b_1 b_4^* - b_2 b_3^*)$      \\

$C_{z^\prime}$ $d\sigma/dt$ & $|S_2|^2-|S_1|^2-|N|^2+|D|^2$ &$-2 Re(b_1 b_4^* + b_2 b_3^*)$     \\

\hline
\end{tabular}
\label{table:amps}
\end{table*}

Barker {\it et al.}~\cite{barker} discuss which combinations of
measured observables lead to complete determination of the amplitudes
free of discrete ambiguities.  In addition to the four measurements
$A$, $\Sigma$, $T$, and $P$ they found that five double-spin
observables were needed, with no four of them coming from the same set
of Beam-Target, Beam-Recoil, or Target-Recoil observables. Chiang and
Tabakin showed~\cite{tabakin1}, however, that with careful selection
of observables, a full determination of the amplitudes is possible
with only four double polarization observable measurements.  Still,
this calls for a far-reaching program to measure the three single spin
observables and at least four double spin observables chosen correctly
from the available 12.  According to the results in
Ref.~\cite{tabakin1}, the present measurements of $C_x$ and $C_z$ can
be combined with almost any other pair of double spin observables to
attain the desired full separation.

At present the only well-measured quantities for hyperons are the
cross sections~\cite{bradforddsdo, bonn2}, induced recoil polarization
$P$~\cite{mcnabb, bonn2, graal}, beam asymmetry $\Sigma$~\cite{leps},
and the present results for $C_x$ and $C_z$.  In the future, CLAS
results are expected for $\Sigma$, $O_x$, $O_z$, and, pending the
operation of a suitable polarized target~\cite{frost}, all the
remaining double-spin observables.  Thus, one cannot expect the
present set of measurements to uniquely specify any of the underlying
production amplitudes, but manipulation of the expressions in
Table~\ref{table:amps} reveals how much is accessible, in principle,
from the information available with these new results.  In the
transversity representation, for example, let $b_i = r_i e^{-i\phi_i}$
and let $A$ represent the reduced cross section.  Then one sees
immediately that
\begin{equation}
A+P = 2(r_1^2 + r_3^2)
\end{equation}
\begin{equation}
A-P = 2(r_2^2 + r_4^2),
\end{equation}
and after some algebra we find
\begin{eqnarray}
\frac{C_{z^\prime}^2 - C_{x^\prime}^2}{C_{z^\prime}^2 + C_{x^\prime}^2} 
= \cos 2(\phi_2 - \phi_3) = \cos 2(\phi_4 - \phi_1).
\end{eqnarray}
The latter statement is true if we select
\begin{equation}
(\phi_1 + \phi_2) - (\phi_3 + \phi_4) \equiv 0
\end{equation}
to fix the overall phase.  From present results, one 
thus obtains only the sums of squared magnitudes of pairs of
amplitudes,  and the
difference between two pairs of phases.
Similar  expressions are obtained in the helicity
representation.  Thus, while a few constraints are placed on the
amplitudes by these measurements, more information is needed to make the
measurements a ``complete'' set.

\section{\label{sec:conclusions}CONCLUSIONS}

In summary, we have presented results from an experimental
investigation of the beam-recoil polarization observables $C_x$ and
$C_z$ for $\Lambda$ and $\Sigma^0$ hyperon photoproduction from the
proton, in the energy range from threshold through the nucleon
resonance region.  These are the first measurements of these
observables.  It is notable that the $\hat z$ component of $\Lambda$
polarization transfer is large and positive, indeed near $+1.0$, over
a broad range of kinematics, where $\hat z$ is the direction of the
initial state photon circular polarization.  It is remarkable that the
$\Lambda$ hyperon is produced fully polarized at all values of $W$ and
scattering angle for a fully circularly polarized beam.  The direction
of this polarization is mostly along $\hat z$, but we have shown how
$C_x$ and $P$ also are substantial in some kinematic regions.  This
phenomenon signifies some as yet unidentified dynamics in the
photoproduction of strangeness.  The $\Sigma^0$ hyperon was measured
with lesser precision, but it is clear that it does not exhibit the
same qualitative behavior, which is perhaps not a surprise since the
spin structure of the $\Sigma^0$ and $\Lambda$ are different.  There
are no existing hadrodynamic or Regge models that do a good job of
predicting these results, so it can be expected that reconsideration
of these models in view of these new results may lead to new insights
into the dynamics of strange quark photoproduction.

\begin{acknowledgments}
We thank the staff of the Accelerator and the Physics Divisions at
Thomas Jefferson National Accelerator Facility who made this
experiment possible.  We thank J. Soffer and A. Afanasev for useful
discussions.  This work was supported in part by the Istituto
Nazionale di Fisica Nucleare, the French Centre National de la
Recherche Scientifique, the French Commissariat \`{a} l'Energie
Atomique, the U.S. Department of Energy, the National Science
Foundation, an Emmy Noether grant from the Deutsche
Forschungsgemeinschaft and the Korean Science and Engineering
Foundation.  The Southeastern Universities Research Association (SURA)
operated Jefferson Lab under United States DOE contract
DE-AC05-84150 during this work.
\end{acknowledgments}


\appendix 
\section{Proton Angular Distribution in the $\Sigma^0$ Rest Frame}
\label{app:sigmadecay}

We compute the angular distribution of protons resulting from the
decay of polarized $\Sigma^0$ ground state hyperons in the $\Sigma^0$
rest frame.  The $\Sigma^0$ hyperon decays 100\% according to
\begin{equation}
\Sigma^0 \to \gamma + \Lambda 
\end{equation}
and the $\Lambda$ decays with a 64\% branch via
\begin{equation}
\Lambda \to \pi^- + p.
\end{equation}
A $\Sigma^0$ produced in a given reaction will generally be polarized
to some degree, $\vec P_{\Sigma^0}$, and the $\Lambda$ arising in the
decay will preserve part of the polarization.  In the rest frame of
the $\Lambda$ hyperon, we have the well-known parity-violating mesonic
weak decay asymmetry that allows measurement of the polarization of
the $\Lambda$ hyperon.  For $\Lambda$ polarization component,
$P_{\Lambda i}$, along a given axis in space, where $i \in \{x,y,z\}$,
the proton intensity distribution, $I(\cos \theta_{p i})$, as a
function of polar angle $\theta_{p i}$ is given by
\begin{equation}
I(\cos \theta_{p i}) = \textstyle\frac{1}{2}(1+\alpha P_\Lambda \cos \theta_{p i}),
\end{equation}
where the value of the weak decay asymmetry parameter, $\alpha$, is
0.642 \cite{pdg06}.  This phenomenon arises from the interference of
the parity violating $S$ and parity conserving $P$ -wave decay
amplitudes~\cite{perkins}.  To determine the $\Lambda$ polarization
component, $P_{\Lambda i}$, one computes the distribution of protons
with respect to $\cos \theta_{p i}$, and then determines the slope of
the resulting straight line that is proportional to $P_{\Lambda i}$.
This procedure must be performed in the $\Lambda$ rest frame.

\begin{figure}[ht]
\centering
\includegraphics[scale=0.35,angle=-90.0]{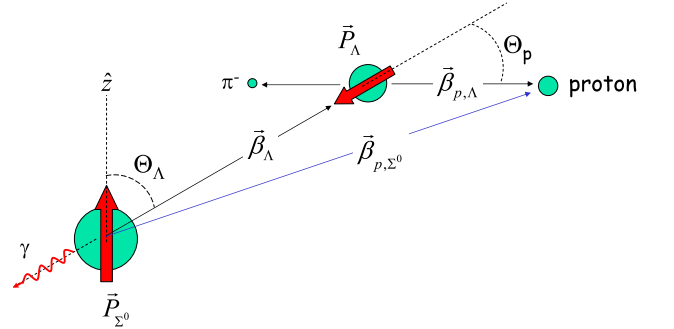}
\caption[]{(Color online) 
The $\Sigma^0$ hyperon, polarized along the $\hat{z}$ axis, decays to
a $\gamma$ and a $\Lambda$ at some angle $\theta_\Lambda$.  The
$\Lambda$ is polarized as shown, traveling at speed
$\beta_\Lambda$. In the $\Lambda$ rest frame, the $\Lambda$ decays into a
$\pi^-$ and a proton, where the proton emission angle with respect to
$\hat{\beta}_\Lambda$ is $\theta_p$ and the speed of the proton 
is $\beta_{p,\Lambda}$. 
  }
\label{fig:a}
\end{figure}

In the rest frame of a $\Sigma^0$ hyperon the first decay is always a
magnetic dipole $(M1)$ transition to a photon and a $\Lambda$.  The
$\Sigma^0$, with $J^\pi = 1/2^+$, decays to a $\Lambda$ with $J^\pi =
1/2^+$, and a photon with $J^\pi = 1^-$.  This is shown schematically
in Fig.~\ref{fig:a}.  As discussed below, for a given $\Sigma^0$
polarization axis it can be shown that the angular distribution of
this decay is isotropic in the decay angle $\theta_\Lambda$.  Crucial
for this discussion is that the decay $\Lambda$ is polarized in an
angle-dependent way.  If the parent $\Sigma^0$ has polarization
$\vec{P}_{\Sigma^0}$, then the daughter $\Lambda$ has polarization
$\vec{P}_\Lambda$ given by
\begin{equation}
\vec{P}_\Lambda(\theta_\Lambda) = -|\vec{P}_{\Sigma^0}|
(\hat{z}\cdot\hat{\beta}_\Lambda)\hat{\beta}_\Lambda,
\end{equation}
where $\vec{\beta}_\Lambda$ is the velocity vector of the $\Lambda$ in
the $\Sigma^0$ rest frame.  This relationship arises from evaluating
the expectation value of the spin operator of the $\Lambda$ in terms
of the transition matrix for this electromagnetic decay~\cite{gatto,
dreitlein}.  This equation says that the $\Lambda$ is polarized along
the axis it is emitted, with its magnitude scaled by the cosine of the
emission angle, $\theta_\Lambda$, as indicated in the figure.

In the $\Lambda$ rest frame, then, the decay angular distribution of
the protons can be written
\begin{equation}
I(\hat{\beta}_{p,\Lambda}) = c(1-\alpha
 P_{\Sigma^0}(\hat{z}\cdot\hat{\beta}_\Lambda)
(\hat{\beta}_\Lambda\cdot\hat{\beta}_{p,\Lambda})),
\end{equation}
where $c$ is a normalization constant, or equivalently as
\begin{equation}
I(\cos \theta_p) = c(1-\alpha
 P_{\Sigma^0} \cos \theta_\Lambda \cos \theta_p).
\label{eq:angdist}
\end{equation}
In situations where the photons are not detected, and the acceptance
for the $\Lambda$ decay products is taken into account properly, we
can integrate over all values of $\theta_\Lambda$.  The only 
direction along which to measure an asymmetry is then $\hat{z}$, and
we must measure the proton angle from this axis, which we will call
$\theta_{p,\Sigma^0}$; this projection introduces another factor of
$\cos \theta_\Lambda$.  The solid-angle weighted average of
$\cos^2 \theta_\Lambda$ is $1/3$, leading to the equation
\begin{equation}
I_{avg}(\theta_{p,\Sigma^0}) = c(1-\alpha P_{\Sigma^0} \frac{1}{3}
 \cos \theta_{p,\Sigma^0})
\end{equation}
for the average distribution of protons in the $\Lambda$ hyperon rest
frame.  Thus, if the direction of the $\Lambda$ is not explicitly
measured, the effective polarization component of the $\Lambda$
reduces along $\hat z$ to the relationship
\begin{equation}
P_\Lambda = -\frac{1}{3}P_{\Sigma^0}.
\end{equation}
As a mnemonic, one can say that the average $\Lambda$ polarization is
$-1/3$ of the $\Sigma^0$ polarization.  However, this statement is
true only in the sense of averaging over all possible $\Lambda$
emission angles.

Now we reach the statement of the problem at hand: what is the angular
distribution of the protons from the decay of the $\Lambda$'s when
measured in the $\Sigma^0$ rest frame instead of the $\Lambda$ rest
frame?  That is, how can the polarization of the parent $\Sigma^0$ be
determined without boosting the protons to the rest frame of the
$\Lambda$?  This problem arises, for example, in the case of the
fixed-target reaction
\begin{eqnarray}
\gamma + p &\to& K^+ + \vec\Sigma^0\nonumber\\ &\to&
           K^+ + (\gamma) + \vec\Lambda \\ &\to& K^+ +
           (\gamma)+ (\pi^-) + p, \nonumber
\end{eqnarray}
where the particles in parentheses are not detected and the vectors
designate the polarized hyperons.  The photon and the kaon define the
boost to the $\Sigma^0$ rest frame, but without detecting the $\gamma$
or the $\pi^-$ it is impossible to define the boost to the $\Lambda$
rest frame.  Determination of the induced or transferred polarizations
of the $\Sigma^0$ necessitates using the angular distribution of the
protons in the $\Sigma^0$ frame.  There is enough kinematic definition
to boost the detected proton to the $\Sigma^0$ rest frame, hence we
need to compute the expected angular distribution of the protons in
that frame.

\subsection{The Calculations}

The polarization of the parent $\Sigma^0$ particle is the expectation
value of the Pauli spin operator, $P_{\Sigma^0} = <\vec
\sigma>_\Sigma$.  In a basis where the initial polarization direction
is the quantization axis, 
the $\Lambda$ spin either is flipped or is not
flipped relative to the $\Sigma^0$ spin.  If the parent particle is in
the $m_\Sigma = +1/2$ state, then it can be shown that the non-spin
flip transition leads to an angular distribution, $I_{+1/2}$,
proportional to $(1-\cos^2\theta_\Lambda)$.  The angular distribution
for spin flip, $I_{-1/2}$, is proportional to
$(1+\cos^2\theta_\Lambda)$.  Summing these two equal-strength
non-interfering final states leads to two predictions.  First, the net
angular distribution of the $\Lambda$'s in the $\Sigma^0$ rest frame
is isotropic, namely
\begin{equation}
I(\theta_\Lambda) \sim I_{+1/2} + I_{-1/2} \sim 1.
\end{equation}
Second, the polarization of the $\Lambda$ hyperons is given by
\begin{equation}
P_\Lambda(\theta_\Lambda) = P_{\Sigma^0} \frac{I_{+1/2} -
I_{-1/2}}{I_{+1/2} + I_{-1/2}} = -P_{\Sigma^0} \cos^2{\theta_\Lambda}
\end{equation}
as stated in the introduction.  Integration of the
$I_m(\theta_\Lambda)$'s over all values of $\theta_\Lambda$ leads to
the result that 1/3 of the time the transition does not flip the spin
({\it i.e.}  $m_\Lambda = +1/2$), while 2/3 of the time the transition
flips the spin $(m_\Lambda =-1/2)$.  The net average polarization of
the $\Lambda$ along the initial polarization axis is then $-1/3$ of
the parent $\Sigma^0$ polarization.  We have done the detailed
calculation of these results ourselves, and found corroboration in
several places~\cite{gatto,dreitlein,tabakin2}.  However, the
calculation of the proton distribution in the $\Sigma^0$ rest frame
requires additional considerations.

In the $\Sigma^0$ rest frame the $\Lambda$ and $\gamma$ are produced
with a momentum of 74.48 MeV/c, which corresponds to a speed of the
$\Lambda$ of $\beta_\Lambda = 0.0666$.  In the $\Lambda$ rest frame
the proton and the $\pi^-$ are produced with a momentum of 100.58
MeV/c, which corresponds to a speed of the proton of $\beta_p =
0.1072$.  Thus, both the $\Lambda$ and the proton are non-relativistic
in the $\Sigma^0$ rest frame, so we will treat the frame
transformation in terms of simple non-relativistic velocity addition.
That is, we compute a weighted average over all possible $\Lambda$
velocities in the $\Sigma^0$ frame, $\vec \beta_\Lambda$, and all
proton velocities in the $\Lambda$ frame, $\vec \beta_{p,\Lambda}$:
\begin{equation}
\vec \beta_{p,\Sigma^0} = \vec \beta_{p,\Lambda} + \vec \beta_\Lambda.
\end{equation}
This can be computed either with an explicit numerical integration or
by integration using a Monte Carlo technique.

\subsubsection{Explicit Integration}

To compute the proton distribution in the $\Sigma^0$ rest frame,
$I_p(\cos \theta_{p,{\Sigma^0}})$, by means of an integration over all
possible proton and $\Lambda$ orientations, each angle combination
must be properly weighted by the underlying intensity
distribution and the proper differential area element.  As discussed
above, the decay-$\Lambda$ distribution is isotropic, and so the
density in three dimensions is equal to $1/4\pi$.  The proton
distribution in the $\Lambda$ rest frame in three dimensions is given
by $(1/2\pi)I(\theta_p)$, where $I(\theta_p)$ is given by Eq.~\ref{eq:angdist}.

We take the initial polarization, $P_{\Sigma^0}$, to be 100\%. 
The complete expression for evaluating the proton angular
distribution in the $\Sigma^0$ rest frame is

\begin{widetext}
\begin{eqnarray}
I_p(\cos \theta_{p,{\Sigma^0}}) = \int_{\theta_\Lambda=0}^{\pi}
\frac{1}{4\pi} \int_{\theta_p=0}^{\pi} \int_{\phi_p=0}^{2\pi}
\delta(\vec{\beta}_{p,\Sigma^0}(\theta_{p,\Sigma^0}) - ( \vec
{\beta}_\Lambda(\theta_\Lambda) + \vec{\beta}_{p,\Lambda}(\theta_p,\phi_p))
\nonumber\\ 
\times\textstyle\frac{1}{2} \left( 1+ \alpha(-\cos^2\theta_\Lambda)\cos\theta_p
\right) d\phi_p \sin\theta_p d\theta_p
\sin\theta_\Lambda d\theta_\Lambda.
\end{eqnarray}
\end{widetext}
The delta function formally enforces the requirement of selecting all
those vector combinations of velocities which lead to a given value of
the proton angle in the $\Sigma^0$ rest frame.  In practice, the
integral was evaluated by numerically sweeping over all values of
$\theta_{p}$, $\phi_{p}$, and $\theta_{\Lambda}$, and accumulating the
distribution of proton angles in the $\Sigma^0$ rest frame,
$\theta_{p,\Sigma^0}$, with the weighting given by the rest of the
integrand.

\begin{figure}[ht]
\centering
\includegraphics[scale=0.35]{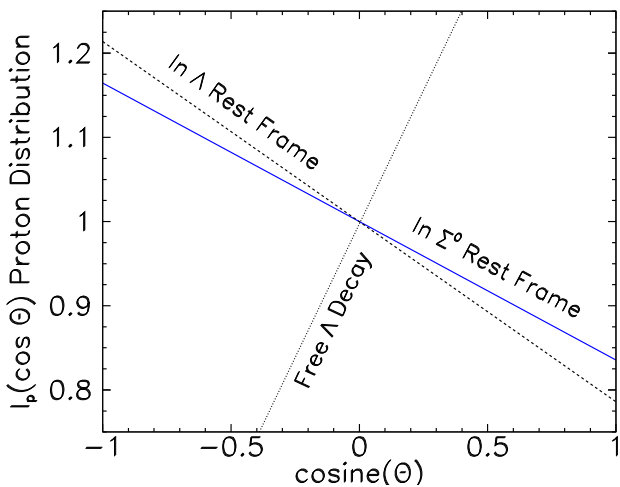}
\caption[]{(Color online) 
The solid (blue) line shows the proton angular distribution in the
$\Sigma^0$ rest frame, where $\theta$ is the polar angle with respect
to the polarization axis $\hat{z}$.  
The dashed (black) line
shows the expected slope of the proton distribution in the $\Lambda$
rest frame, where the angle $\theta$ in the graph is construed as the
proton angle $\theta_{p}$, {\it i.e.} measured from the axis of the
$\Lambda$'s velocity.  The dotted (black) curve shows,
for reference, the slope of the proton distribution in the case of
fully polarized $\Lambda$ hyperons decaying, where the angle is
$\theta_{p}$ with respect to the $\Lambda$ polarization axis.
  }
\label{fig:b}
\end{figure}

The result is shown in Fig.~\ref{fig:b}.  The calculation assumes a
fully polarized $\Sigma^0$ hyperon.  The solid line shows the result
of the integration.  In effect, the straight-line proton distribution
in the $\Lambda$ rest frame (dot-dashed line) is shifted by the
transformation to the $\Sigma^0$ rest frame.  The fact that this
result is a straight line rather than some inflected curve is
significant.  It shows that the $\Sigma^0$ polarization can be
determined using the same method, in essence, as when determining a
$\Lambda$'s polarization.  Experimental data can be fitted with this
slope and the actual polarization of the parent particle can be
deduced from the scale factor.  The first moment of the calculated
distribution gives the slope.  The value is $-0.1646$ in the
$\Sigma^0$ rest frame.  In the rest frame of the $\Lambda$, when all 
possible decay-$\Lambda$ angles are averaged the slope
is given by $-(1/3)\alpha = -0.214$.  Thus, the slope of the asymmetry
is reduced by the frame transformation by an amount given by $
1/(.2140/.1646) = 1/1.300 = 0.769$. Thus, one can say the frame
transformation reduces the slope by 30.0\%, or alternatively, that the
effective weak decay constant, $\alpha_{eff}$ is
\begin{equation}
\alpha_{eff} = -\alpha \times 0.256 = \alpha\times\left (-\frac{1}{3.90}\right ).
\end{equation}

\subsubsection{Monte Carlo Simulation}

Two separate three-dimensional Monte Carlo simulations of the problem
were performed. The frame-transformation calculation was treated
non-relativistically, as in the explicit integration discussed in the
previous section.  The difference in approach entailed random weighted
selection of the decay directions at each step, which eliminated the
need to separately compute the solid-angle weighting factors.  The
results of the Monte Carlo and of the direct numerical integration
methods agreed to three significant figures.

\subsection{Appendix Summary}

Using two independent calculation methods we have numerically
evaluated the angular distribution of protons that arise from the
two-step decay of $\Sigma^0$ hyperons in their rest frame.  The result
is a decay asymmetry that is well represented by a constant slope in
$\cos{\theta_{p,{\Sigma^0}}}$.  The distribution has a slope that is
reduced by 30.0\% with respect to the average slope expected in the
rest frame of the intermediate $\Lambda$ hyperon.  The effective weak
decay constant is $-0.165$.



\section{Numerical Data}
\label{app:data}

The polarization transfer results from the present work are given
below.  Each row gives the values for $C_x$ and $C_z$ for the stated
values of photon energy and $\cos\theta_{K^+}^{c.m.}$, where
$\theta_{K^+}^{c.m.}$ is the center-of-mass angle of the kaon.  The
quoted uncertainties are the statistical errors resulting from the
proton yield asymmetry fitting combined with the point-to-point
systematic uncertainty in the fitting procedures.  Global systematic
uncertainties were discussed in the main text.  A zero value for an
uncertainty means that no data point was extracted at that energy and
angle.  Electronic tabulations of the results are available from
several archival sources~\cite{bradfordthesis}, \cite{clasdb},
\cite{durham}, \cite{contact}.

\begingroup
\begin{longtable}{cccccccc}

\caption{ Results of CLAS measurements of $\gamma+p \to K^+ +
\Lambda$.  The columns marked $C_x$ and $C_z$ are the polarization
transfer coefficients of the photon to the hyperon in the $\gamma p$
center-of-mass (c.m.)  frame, where $\hat{x}$ and $\hat{z}$ are
defined in the text.  The column $\cos\theta_{K^+}^{c.m.}$ gives the
c.m. angle of the produced $K^+$ meson.  The columns marked $\delta
C_x$ and $\delta C_z$ are the associated uncertainties.
\label{tab:table1}
}\\

\hline\hline
Index  & $E_\gamma$  & W & $\cos\theta_K^{c.m.}$ 
& $C_x$  & $\delta {C_x}$ & $C_z$  & $\delta C_z$ \\  
& (GeV) & (GeV) \\
\hline
\endfirsthead

\multicolumn{8}{c}%
{{\bfseries \tablename\ \thetable{} -- continued}} \\

\hline\hline
Index  & $E_\gamma$  & W & $\cos\theta_K^{c.m.}$ 
& $C_x$  & $\delta {C_x}$ & $C_z$  & $\delta C_z$ \\  
& (GeV) & (GeV) \\
\hline
\endhead

\hline \hline
\endlastfoot

\input{tablelambda}

\end{longtable}
\endgroup



\begingroup
\begin{longtable}{cccccccc}

\caption{ Results of CLAS measurements of $\gamma+p \to K^+ +
\Sigma^0$.  The columns marked $C_x$ and $C_z$ are the polarization
transfer coefficients of the photon to the hyperon in the $\gamma p$
center-of-mass (c.m.)  frame, where $\hat{x}$ and $\hat{z}$ are
defined in the text.  The column $\cos\theta_{K^+}^{c.m.}$ gives the
c.m. angle of the produced $K^+$ meson.  The columns marked $\delta
C_x$ and $\delta C_z$ are the associated uncertainties.
\label{tab:table2}
}\\

\hline\hline
Index  & $E_\gamma$  & W & $\cos\theta_K^{c.m.}$ 
& $C_x$  & $\delta {C_x}$ & $C_z$  & $\delta C_z$ \\  
& (GeV) & (GeV) \\
\hline
\endfirsthead

\multicolumn{8}{c}%
{{\bfseries \tablename\ \thetable{} -- continued}} \\

\hline\hline
Index  & $E_\gamma$  & W & $\cos\theta_K^{c.m.}$ 
& $C_x$  & $\delta {C_x}$ & $C_z$  & $\delta C_z$ \\  
& (GeV) & (GeV) \\
\hline
\endhead

\hline \hline
\endlastfoot

\input{tablesigma0}

\end{longtable}
\endgroup

\end{document}

%% file: authors.tex
%
%
%
%

\newcommand*{\CMU}{Carnegie Mellon University, Pittsburgh, Pennsylvania 15213}
\affiliation{\CMU}
\newcommand*{\ANL}{Argonne National Laboratory, Argonne, Illinois 60439}
\affiliation{\ANL}
\newcommand*{\ASU}{Arizona State University, Tempe, Arizona 85287-1504}
\affiliation{\ASU}
\newcommand*{\UCLA}{University of California at Los Angeles, Los Angeles, California  90095-1547}
\affiliation{\UCLA}
\newcommand*{\CSU}{California State University, Dominguez Hills, Carson, CA 90747}
\affiliation{\CSU}
\newcommand*{\CUA}{Catholic University of America, Washington, D.C. 20064}
\affiliation{\CUA}
\newcommand*{\SACLAY}{CEA-Saclay, Service de Physique Nucl\'eaire, 91191 Gif-sur-Yvette, France}
\affiliation{\SACLAY}
\newcommand*{\CNU}{Christopher Newport University, Newport News, Virginia 23606}
\affiliation{\CNU}
\newcommand*{\UCONN}{University of Connecticut, Storrs, Connecticut 06269}
\affiliation{\UCONN}
\newcommand*{\DUKE}{Duke University, Durham, North Carolina 27708-0305}
\affiliation{\DUKE}
\newcommand*{\ECOSSEE}{Edinburgh University, Edinburgh EH9 3JZ, United Kingdom}
\affiliation{\ECOSSEE}
\newcommand*{\FU}{Fairfield University, Fairfield CT 06824}
\affiliation{\FU}
\newcommand*{\FIU}{Florida International University, Miami, Florida 33199}
\affiliation{\FIU}
\newcommand*{\FSU}{Florida State University, Tallahassee, Florida 32306}
\affiliation{\FSU}
\newcommand*{\GWU}{The George Washington University, Washington, DC 20052}
\affiliation{\GWU}
\newcommand*{\ECOSSEG}{University of Glasgow, Glasgow G12 8QQ, United Kingdom}
\affiliation{\ECOSSEG}
\newcommand*{\ISU}{Idaho State University, Pocatello, Idaho 83209}
\affiliation{\ISU}
\newcommand*{\INFNFR}{INFN, Laboratori Nazionali di Frascati, 00044 Frascati, Italy}
\affiliation{\INFNFR}
\newcommand*{\INFNGE}{INFN, Sezione di Genova, 16146 Genova, Italy}
\affiliation{\INFNGE}
\newcommand*{\ORSAY}{Institut de Physique Nucleaire ORSAY, Orsay, France}
\affiliation{\ORSAY}
\newcommand*{\ITEP}{Institute of Theoretical and Experimental Physics, Moscow, 117259, Russia}
\affiliation{\ITEP}
\newcommand*{\JMU}{James Madison University, Harrisonburg, Virginia 22807}
\affiliation{\JMU}
\newcommand*{\KYUNGPOOK}{Kyungpook National University, Daegu 702-701, South Korea}
\affiliation{\KYUNGPOOK}
\newcommand*{\MIT}{Massachusetts Institute of Technology, Cambridge, Massachusetts  02139-4307}
\affiliation{\MIT}
\newcommand*{\UMASS}{University of Massachusetts, Amherst, Massachusetts  01003}
\affiliation{\UMASS}
\newcommand*{\MOSCOW}{Moscow State University, General Nuclear Physics Institute, 119899 Moscow, Russia}
\affiliation{\MOSCOW}
\newcommand*{\UNH}{University of New Hampshire, Durham, New Hampshire 03824-3568}
\affiliation{\UNH}
\newcommand*{\NSU}{Norfolk State University, Norfolk, Virginia 23504}
\affiliation{\NSU}
\newcommand*{\OHIOU}{Ohio University, Athens, Ohio  45701}
\affiliation{\OHIOU}
\newcommand*{\ODU}{Old Dominion University, Norfolk, Virginia 23529}
\affiliation{\ODU}
\newcommand*{\PITT}{University of Pittsburgh, Pittsburgh, Pennsylvania 15260}
\affiliation{\PITT}
\newcommand*{\RPI}{Rensselaer Polytechnic Institute, Troy, New York 12180-3590}
\affiliation{\RPI}
\newcommand*{\RICE}{Rice University, Houston, Texas 77005-1892}
\affiliation{\RICE}
\newcommand*{\URICH}{University of Richmond, Richmond, Virginia 23173}
\affiliation{\URICH}
\newcommand*{\SCAROLINA}{University of South Carolina, Columbia, South Carolina 29208}
\affiliation{\SCAROLINA}
\newcommand*{\JLAB}{Thomas Jefferson National Accelerator Facility, Newport News, Virginia 23606}
\affiliation{\JLAB}
\newcommand*{\UNIONC}{Union College, Schenectady, NY 12308}
\affiliation{\UNIONC}
\newcommand*{\VT}{Virginia Polytechnic Institute and State University, Blacksburg, Virginia   24061-0435}
\affiliation{\VT}
\newcommand*{\VIRGINIA}{University of Virginia, Charlottesville, Virginia 22901}
\affiliation{\VIRGINIA}
\newcommand*{\WM}{College of William and Mary, Williamsburg, Virginia 23187-8795}
\affiliation{\WM}
\newcommand*{\YEREVAN}{Yerevan Physics Institute, 375036 Yerevan, Armenia}
\affiliation{\YEREVAN}
\newcommand*{\deceased}{Deceased}
\newcommand*{\NOWOHIOU}{Ohio University, Athens, Ohio  45701}
\newcommand*{\NOWINDSTRA}{Systems Planning and Analysis, Alexandria, Virginia 22311}
\newcommand*{\NOWUNH}{University of New Hampshire, Durham, New Hampshire 03824-3568}
\newcommand*{\NOWCUA}{Catholic University of America, Washington, D.C. 20064}
\newcommand*{\NOWSANPAULO}{San Paulo University, Brazil}
\newcommand*{\NOWNONE}{unknown, }
\newcommand*{\NOWUMASS}{University of Massachusetts, Amherst, Massachusetts  01003}
\newcommand*{\NOWMIT}{Massachusetts Institute of Technology, Cambridge, Massachusetts  02139-4307}
\newcommand*{\NOWECOSSEE}{Edinburgh University, Edinburgh EH9 3JZ, United Kingdom}
\newcommand*{\NOWBONN}{Helmholtz-Institut f\"ur Strahlen- und Kernphysik, Nussallee 14-16, 53115 Bonn, Germany}
\newcommand*{\NOWROCH}{University of Rochester, Rochester, NY 14627}

\author {R.~Bradford} 
\altaffiliation[Current address: ]{\NOWROCH}
\affiliation{\CMU}
\author {R.A.~Schumacher} 
\affiliation{\CMU}
\author {G.~Adams} 
\affiliation{\RPI}
\author {M.J.~Amaryan} 
\affiliation{\ODU}
\author {P.~Ambrozewicz} 
\affiliation{\FIU}
\author {E.~Anciant} 
\affiliation{\SACLAY}
\author {M.~Anghinolfi} 
\affiliation{\INFNGE}
\author {B.~Asavapibhop} 
\affiliation{\UMASS}
\author {G.~Asryan} 
\affiliation{\YEREVAN}
\author {G.~Audit} 
\affiliation{\SACLAY}
% ...
\collaboration{The CLAS Collaboration}
     \noaffiliation

%% file: tablelambda.tex
    1) & 1.032 & 1.679 & -0.75 &  0.000 &  0.000 &   0.000 &  0.000 \\
    2) & 1.032 & 1.679 & -0.55 &  0.114 &  0.724 &   0.698 &  0.542 \\
    3) & 1.032 & 1.679 & -0.35 & -0.960 &  0.269 &   0.686 &  0.230 \\
    4) & 1.032 & 1.679 & -0.15 & -0.304 &  0.186 &   0.812 &  0.172 \\
    5) & 1.032 & 1.679 &  0.05 & -0.470 &  0.169 &   0.917 &  0.160 \\
    6) & 1.032 & 1.679 &  0.25 & -0.700 &  0.175 &   0.838 &  0.154 \\
    7) & 1.032 & 1.679 &  0.45 & -0.444 &  0.139 &   0.555 &  0.131 \\
    8) & 1.032 & 1.679 &  0.65 & -0.216 &  0.154 &   0.821 &  0.135 \\
    9) & 1.032 & 1.679 &  0.85 & -0.126 &  0.190 &   0.901 &  0.189 \\
   10) & 1.132 & 1.734 & -0.75 &  0.000 &  0.000 &   0.000 &  0.000 \\
   11) & 1.132 & 1.734 & -0.55 & -0.453 &  0.310 &   0.451 &  0.280 \\
   12) & 1.132 & 1.734 & -0.35 & -0.543 &  0.186 &   0.742 &  0.179 \\
   13) & 1.132 & 1.734 & -0.15 & -0.260 &  0.136 &   0.940 &  0.135 \\
   14) & 1.132 & 1.734 &  0.05 & -0.301 &  0.123 &   1.004 &  0.117 \\
   15) & 1.132 & 1.734 &  0.25 & -0.426 &  0.108 &   0.983 &  0.110 \\
   16) & 1.132 & 1.734 &  0.45 & -0.235 &  0.089 &   0.752 &  0.085 \\
   17) & 1.132 & 1.734 &  0.65 & -0.189 &  0.103 &   0.893 &  0.084 \\
   18) & 1.132 & 1.734 &  0.85 & -0.262 &  0.160 &   1.017 &  0.111 \\
   19) & 1.232 & 1.787 & -0.75 &  0.140 &  1.021 &   0.974 &  0.426 \\
   20) & 1.232 & 1.787 & -0.55 & -0.224 &  0.228 &   1.002 &  0.205 \\
   21) & 1.232 & 1.787 & -0.35 & -0.065 &  0.138 &   0.762 &  0.140 \\
   22) & 1.232 & 1.787 & -0.15 & -0.375 &  0.112 &   0.848 &  0.113 \\
   23) & 1.232 & 1.787 &  0.05 &  0.041 &  0.097 &   0.779 &  0.102 \\
   24) & 1.232 & 1.787 &  0.25 & -0.185 &  0.081 &   0.983 &  0.088 \\
   25) & 1.232 & 1.787 &  0.45 & -0.072 &  0.069 &   0.905 &  0.087 \\
   26) & 1.232 & 1.787 &  0.65 & -0.061 &  0.069 &   1.021 &  0.077 \\
   27) & 1.232 & 1.787 &  0.85 & -0.086 &  0.094 &   1.001 &  0.100 \\
   28) & 1.332 & 1.839 & -0.75 &  0.024 &  0.303 &   0.982 &  0.238 \\
   29) & 1.332 & 1.839 & -0.55 & -0.094 &  0.148 &   0.869 &  0.146 \\
   30) & 1.332 & 1.839 & -0.35 & -0.237 &  0.109 &   1.067 &  0.112 \\
   31) & 1.332 & 1.839 & -0.15 & -0.160 &  0.089 &   1.067 &  0.094 \\
   32) & 1.332 & 1.839 &  0.05 & -0.056 &  0.081 &   0.891 &  0.098 \\
   33) & 1.332 & 1.839 &  0.25 & -0.086 &  0.067 &   0.943 &  0.074 \\
   34) & 1.332 & 1.839 &  0.45 & -0.139 &  0.063 &   1.016 &  0.066 \\
   35) & 1.332 & 1.839 &  0.65 & -0.044 &  0.062 &   0.998 &  0.064 \\
   36) & 1.332 & 1.839 &  0.85 &  0.015 &  0.081 &   0.998 &  0.080 \\
   37) & 1.433 & 1.889 & -0.75 & -0.099 &  0.155 &   1.125 &  0.149 \\
   38) & 1.433 & 1.889 & -0.55 & -0.266 &  0.106 &   0.954 &  0.117 \\
   39) & 1.433 & 1.889 & -0.35 & -0.145 &  0.097 &   1.203 &  0.101 \\
   40) & 1.433 & 1.889 & -0.15 &  0.050 &  0.078 &   1.044 &  0.086 \\
   41) & 1.433 & 1.889 &  0.05 & -0.074 &  0.070 &   0.900 &  0.080 \\
   42) & 1.433 & 1.889 &  0.25 & -0.047 &  0.058 &   1.076 &  0.069 \\
   43) & 1.433 & 1.889 &  0.45 & -0.149 &  0.053 &   0.881 &  0.066 \\
   44) & 1.433 & 1.889 &  0.65 & -0.218 &  0.050 &   0.966 &  0.063 \\
   45) & 1.433 & 1.889 &  0.85 & -0.038 &  0.067 &   1.075 &  0.076 \\
   46) & 1.534 & 1.939 & -0.75 & -0.086 &  0.129 &   0.914 &  0.124 \\
   47) & 1.534 & 1.939 & -0.55 & -0.104 &  0.115 &   0.723 &  0.105 \\
   48) & 1.534 & 1.939 & -0.35 & -0.047 &  0.093 &   1.062 &  0.096 \\
   49) & 1.534 & 1.939 & -0.15 & -0.027 &  0.080 &   0.928 &  0.084 \\
   50) & 1.534 & 1.939 &  0.05 &  0.003 &  0.070 &   0.910 &  0.074 \\
   51) & 1.534 & 1.939 &  0.25 &  0.002 &  0.054 &   0.886 &  0.063 \\
   52) & 1.534 & 1.939 &  0.45 & -0.076 &  0.047 &   0.853 &  0.058 \\
   53) & 1.534 & 1.939 &  0.65 & -0.191 &  0.046 &   1.011 &  0.054 \\
   54) & 1.534 & 1.939 &  0.85 & -0.084 &  0.060 &   1.051 &  0.066 \\
   55) & 1.635 & 1.987 & -0.75 & -0.216 &  0.155 &   0.554 &  0.141 \\
   56) & 1.635 & 1.987 & -0.55 & -0.058 &  0.135 &   0.632 &  0.129 \\
   57) & 1.635 & 1.987 & -0.35 & -0.140 &  0.116 &   0.731 &  0.123 \\
   58) & 1.635 & 1.987 & -0.15 & -0.014 &  0.112 &   0.897 &  0.097 \\
   59) & 1.635 & 1.987 &  0.05 &  0.226 &  0.080 &   1.001 &  0.078 \\
   60) & 1.635 & 1.987 &  0.25 & -0.105 &  0.059 &   0.829 &  0.062 \\
   61) & 1.635 & 1.987 &  0.45 & -0.128 &  0.054 &   0.925 &  0.053 \\
   62) & 1.635 & 1.987 &  0.65 & -0.319 &  0.049 &   0.941 &  0.054 \\
   63) & 1.635 & 1.987 &  0.85 & -0.195 &  0.073 &   0.893 &  0.107 \\
   64) & 1.737 & 2.035 & -0.75 & -0.121 &  0.145 &   0.472 &  0.136 \\
   65) & 1.737 & 2.035 & -0.55 & -0.497 &  0.149 &   0.229 &  0.149 \\
   66) & 1.737 & 2.035 & -0.35 & -0.305 &  0.141 &   0.554 &  0.163 \\
   67) & 1.737 & 2.035 & -0.15 & -0.168 &  0.110 &   0.608 &  0.115 \\
   68) & 1.737 & 2.035 &  0.05 & -0.010 &  0.081 &   0.801 &  0.089 \\
   69) & 1.737 & 2.035 &  0.25 &  0.120 &  0.063 &   0.843 &  0.064 \\
   70) & 1.737 & 2.035 &  0.45 & -0.112 &  0.050 &   1.022 &  0.057 \\
   71) & 1.737 & 2.035 &  0.65 & -0.241 &  0.055 &   0.886 &  0.047 \\
   72) & 1.737 & 2.035 &  0.85 & -0.331 &  0.071 &   0.942 &  0.090 \\
   73) & 1.838 & 2.081 & -0.75 & -0.384 &  0.180 &   0.422 &  0.174 \\
   74) & 1.838 & 2.081 & -0.55 & -0.618 &  0.189 &   0.207 &  0.192 \\
   75) & 1.838 & 2.081 & -0.35 & -0.960 &  0.190 &   0.469 &  0.189 \\
   76) & 1.838 & 2.081 & -0.15 & -0.056 &  0.139 &   0.489 &  0.150 \\
   77) & 1.838 & 2.081 &  0.05 &  0.229 &  0.107 &   0.989 &  0.112 \\
   78) & 1.838 & 2.081 &  0.25 &  0.087 &  0.071 &   0.850 &  0.080 \\
   79) & 1.838 & 2.081 &  0.45 & -0.109 &  0.056 &   0.946 &  0.068 \\
   80) & 1.838 & 2.081 &  0.65 & -0.335 &  0.065 &   0.831 &  0.103 \\
   81) & 1.838 & 2.081 &  0.85 & -0.448 &  0.074 &   0.870 &  0.063 \\
   82) & 1.939 & 2.126 & -0.75 & -0.522 &  0.182 &   0.778 &  0.174 \\
   83) & 1.939 & 2.126 & -0.55 & -1.082 &  0.181 &   0.499 &  0.177 \\
   84) & 1.939 & 2.126 & -0.35 & -0.957 &  0.174 &   0.352 &  0.205 \\
   85) & 1.939 & 2.126 & -0.15 & -0.558 &  0.150 &   0.786 &  0.149 \\
   86) & 1.939 & 2.126 &  0.05 & -0.034 &  0.111 &   0.743 &  0.125 \\
   87) & 1.939 & 2.126 &  0.25 &  0.290 &  0.077 &   0.922 &  0.083 \\
   88) & 1.939 & 2.126 &  0.45 & -0.154 &  0.110 &   1.048 &  0.272 \\
   89) & 1.939 & 2.126 &  0.65 & -0.328 &  0.166 &   0.726 &  0.239 \\
   90) & 1.939 & 2.126 &  0.85 & -0.475 &  0.063 &   0.691 &  0.068 \\
   91) & 2.039 & 2.170 & -0.75 & -0.501 &  0.186 &   0.469 &  0.171 \\
   92) & 2.039 & 2.170 & -0.55 & -0.962 &  0.161 &   0.533 &  0.175 \\
   93) & 2.039 & 2.170 & -0.35 & -0.896 &  0.168 &   0.486 &  0.220 \\
   94) & 2.039 & 2.170 & -0.15 & -0.121 &  0.161 &   0.621 &  0.175 \\
   95) & 2.039 & 2.170 &  0.05 &  0.053 &  0.121 &   0.908 &  0.141 \\
   96) & 2.039 & 2.170 &  0.25 &  0.078 &  0.088 &   1.045 &  0.092 \\
   97) & 2.039 & 2.170 &  0.45 & -0.047 &  0.068 &   1.045 &  0.196 \\
   98) & 2.039 & 2.170 &  0.65 & -0.431 &  0.062 &   0.834 &  0.061 \\
   99) & 2.039 & 2.170 &  0.85 & -0.552 &  0.061 &   0.655 &  0.071 \\
  100) & 2.139 & 2.212 & -0.75 & -0.983 &  0.252 &   0.987 &  0.204 \\
  101) & 2.139 & 2.212 & -0.55 & -0.800 &  0.187 &   0.760 &  0.197 \\
  102) & 2.139 & 2.212 & -0.35 & -0.711 &  0.273 &   0.569 &  0.198 \\
  103) & 2.139 & 2.212 & -0.15 & -0.170 &  0.180 &   0.425 &  0.231 \\
  104) & 2.139 & 2.212 &  0.05 & -0.145 &  0.131 &   0.858 &  0.158 \\
  105) & 2.139 & 2.212 &  0.25 &  0.034 &  0.106 &   0.886 &  0.106 \\
  106) & 2.139 & 2.212 &  0.45 &  0.020 &  0.084 &   1.042 &  0.099 \\
  107) & 2.139 & 2.212 &  0.65 & -0.299 &  0.056 &   0.761 &  0.074 \\
  108) & 2.139 & 2.212 &  0.85 & -0.455 &  0.062 &   0.676 &  0.074 \\
  109) & 2.240 & 2.255 & -0.75 & -0.690 &  0.226 &   0.674 &  0.228 \\
  110) & 2.240 & 2.255 & -0.55 & -0.466 &  0.268 &   0.962 &  0.231 \\
  111) & 2.240 & 2.255 & -0.35 & -0.931 &  0.252 &   0.872 &  0.278 \\
  112) & 2.240 & 2.255 & -0.15 & -0.504 &  0.229 &   0.365 &  0.229 \\
  113) & 2.240 & 2.255 &  0.05 &  0.206 &  0.194 &   0.576 &  0.213 \\
  114) & 2.240 & 2.255 &  0.25 &  0.245 &  0.136 &   1.080 &  0.138 \\
  115) & 2.240 & 2.255 &  0.45 & -0.183 &  0.098 &   1.061 &  0.130 \\
  116) & 2.240 & 2.255 &  0.65 & -0.466 &  0.077 &   0.698 &  0.081 \\
  117) & 2.240 & 2.255 &  0.85 & -0.421 &  0.075 &   0.576 &  0.087 \\
  118) & 2.341 & 2.296 & -0.75 & -0.357 &  0.384 &   1.015 &  0.296 \\
  119) & 2.341 & 2.296 & -0.55 & -0.232 &  0.290 &   0.487 &  0.372 \\
  120) & 2.341 & 2.296 & -0.35 & -0.354 &  0.261 &   1.266 &  0.307 \\
  121) & 2.341 & 2.296 & -0.15 & -0.241 &  0.320 &   0.502 &  0.326 \\
  122) & 2.341 & 2.296 &  0.05 &  0.280 &  0.299 &   1.016 &  0.322 \\
  123) & 2.341 & 2.296 &  0.25 &  0.636 &  0.194 &   0.779 &  0.189 \\
  124) & 2.341 & 2.296 &  0.45 &  0.032 &  0.130 &   1.147 &  0.128 \\
  125) & 2.341 & 2.296 &  0.65 & -0.492 &  0.083 &   0.638 &  0.099 \\
  126) & 2.341 & 2.296 &  0.85 & -0.450 &  0.087 &   0.610 &  0.121 \\
  127) & 2.443 & 2.338 & -0.75 & -0.790 &  0.385 &   0.907 &  0.402 \\
  128) & 2.443 & 2.338 & -0.55 & -0.697 &  0.422 &   1.412 &  0.416 \\
  129) & 2.443 & 2.338 & -0.35 & -0.253 &  0.402 &   1.025 &  0.472 \\
  130) & 2.443 & 2.338 & -0.15 & -0.521 &  0.366 &   0.932 &  0.429 \\
  131) & 2.443 & 2.338 &  0.05 & -0.097 &  0.278 &   0.866 &  0.317 \\
  132) & 2.443 & 2.338 &  0.25 &  0.107 &  0.231 &   0.499 &  0.293 \\
  133) & 2.443 & 2.338 &  0.45 &  0.191 &  0.156 &   1.439 &  0.170 \\
  134) & 2.443 & 2.338 &  0.65 & -0.393 &  0.334 &   0.508 &  0.386 \\
  135) & 2.443 & 2.338 &  0.85 & -0.416 &  0.157 &   0.360 &  0.172 \\
  136) & 2.543 & 2.377 & -0.75 & -0.393 &  0.432 &   0.396 &  0.410 \\
  137) & 2.543 & 2.377 & -0.55 & -0.007 &  0.420 &   0.281 &  0.669 \\
  138) & 2.543 & 2.377 & -0.35 & -0.938 &  0.466 &   1.102 &  0.369 \\
  139) & 2.543 & 2.377 & -0.15 & -0.188 &  0.406 &   1.170 &  0.471 \\
  140) & 2.543 & 2.377 &  0.05 & -0.521 &  0.403 &   0.525 &  0.596 \\
  141) & 2.543 & 2.377 &  0.25 & -0.289 &  0.401 &   0.390 &  0.290 \\
  142) & 2.543 & 2.377 &  0.45 & -0.426 &  0.189 &   0.809 &  0.235 \\
  143) & 2.543 & 2.377 &  0.65 & -0.258 &  0.126 &   0.698 &  0.145 \\
  144) & 2.543 & 2.377 &  0.85 & -0.450 &  0.114 &   0.504 &  0.164 \\
  145) & 2.642 & 2.416 & -0.75 & -0.135 &  0.569 &   1.640 &  0.497 \\
  146) & 2.642 & 2.416 & -0.55 & -0.102 &  0.554 &   1.580 &  0.624 \\
  147) & 2.642 & 2.416 & -0.35 & -0.903 &  0.472 &   1.385 &  0.367 \\
  148) & 2.642 & 2.416 & -0.15 & -0.866 &  0.533 &   0.649 &  0.478 \\
  149) & 2.642 & 2.416 &  0.05 & -0.052 &  0.534 &   0.587 &  0.459 \\
  150) & 2.642 & 2.416 &  0.25 &  0.122 &  0.405 &  -0.369 &  0.399 \\
  151) & 2.642 & 2.416 &  0.45 & -0.126 &  0.238 &   0.887 &  0.240 \\
  152) & 2.642 & 2.416 &  0.65 & -0.385 &  0.140 &   0.548 &  0.157 \\
  153) & 2.642 & 2.416 &  0.85 & -0.155 &  0.205 &   0.397 &  0.238 \\
  154) & 2.741 & 2.454 & -0.75 &  0.534 &  0.796 &   0.667 &  0.610 \\
  155) & 2.741 & 2.454 & -0.55 &  0.064 &  1.182 &   0.000 &  0.833 \\
  156) & 2.741 & 2.454 & -0.35 &  0.105 &  0.713 &  -0.517 &  1.211 \\
  157) & 2.741 & 2.454 & -0.15 & -0.289 &  0.442 &   1.129 &  0.618 \\
  158) & 2.741 & 2.454 &  0.05 & -0.549 &  0.717 &  -0.171 &  0.480 \\
  159) & 2.741 & 2.454 &  0.25 &  0.071 &  0.402 &   0.117 &  0.354 \\
  160) & 2.741 & 2.454 &  0.45 & -0.083 &  0.233 &   0.906 &  0.318 \\
  161) & 2.741 & 2.454 &  0.65 & -0.474 &  0.213 &   0.809 &  0.222 \\
  162) & 2.741 & 2.454 &  0.85 & -0.449 &  0.212 &   0.283 &  0.527 \\

%% file: tablesigma0.tex
    1) & 1.232 & 1.787 & -0.70 &  0.000 &  0.000 &   0.000 &  0.000 \\
    2) & 1.232 & 1.787 & -0.40 & -1.348 &  0.595 &   0.689 &  0.519 \\
    3) & 1.232 & 1.787 & -0.10 & -0.227 &  0.379 &   1.477 &  0.347 \\
    4) & 1.232 & 1.787 &  0.20 & -0.792 &  0.371 &   0.760 &  0.301 \\
    5) & 1.232 & 1.787 &  0.50 & -0.438 &  0.305 &   1.391 &  0.302 \\
    6) & 1.232 & 1.787 &  0.80 & -0.286 &  0.442 &   1.334 &  0.383 \\
    7) & 1.332 & 1.839 & -0.70 &  0.000 &  0.000 &   0.000 &  0.000 \\
    8) & 1.332 & 1.839 & -0.40 & -1.045 &  0.335 &   0.734 &  0.325 \\
    9) & 1.332 & 1.839 & -0.10 & -0.963 &  0.226 &   1.092 &  0.208 \\
   10) & 1.332 & 1.839 &  0.20 & -0.563 &  0.193 &   1.019 &  0.189 \\
   11) & 1.332 & 1.839 &  0.50 & -0.756 &  0.192 &   0.744 &  0.184 \\
   12) & 1.332 & 1.839 &  0.80 & -0.077 &  0.277 &   0.585 &  0.279 \\
   13) & 1.433 & 1.889 & -0.70 & -0.729 &  0.670 &   0.171 &  0.612 \\
   14) & 1.433 & 1.889 & -0.40 & -0.007 &  0.257 &   0.475 &  0.239 \\
   15) & 1.433 & 1.889 & -0.10 & -0.437 &  0.157 &   0.681 &  0.156 \\
   16) & 1.433 & 1.889 &  0.20 & -0.131 &  0.137 &   0.866 &  0.138 \\
   17) & 1.433 & 1.889 &  0.50 & -0.552 &  0.139 &   0.632 &  0.135 \\
   18) & 1.433 & 1.889 &  0.80 & -0.074 &  0.205 &   0.458 &  0.220 \\
   19) & 1.534 & 1.939 & -0.70 & -1.622 &  0.693 &   1.482 &  0.619 \\
   20) & 1.534 & 1.939 & -0.40 & -0.131 &  0.238 &   0.262 &  0.267 \\
   21) & 1.534 & 1.939 & -0.10 & -0.020 &  0.156 &   1.075 &  0.162 \\
   22) & 1.534 & 1.939 &  0.20 & -0.227 &  0.134 &   0.528 &  0.142 \\
   23) & 1.534 & 1.939 &  0.50 & -0.410 &  0.140 &   0.706 &  0.148 \\
   24) & 1.534 & 1.939 &  0.80 & -0.309 &  0.218 &   0.351 &  0.205 \\
   25) & 1.635 & 1.987 & -0.70 & -0.115 &  0.767 &   0.497 &  0.677 \\
   26) & 1.635 & 1.987 & -0.40 & -0.490 &  0.283 &   0.697 &  0.281 \\
   27) & 1.635 & 1.987 & -0.10 & -0.312 &  0.173 &   0.744 &  0.185 \\
   28) & 1.635 & 1.987 &  0.20 & -0.640 &  0.147 &   1.022 &  0.141 \\
   29) & 1.635 & 1.987 &  0.50 & -0.362 &  0.143 &   0.379 &  0.146 \\
   30) & 1.635 & 1.987 &  0.80 & -0.047 &  0.216 &   0.697 &  0.186 \\
   31) & 1.737 & 2.035 & -0.70 &  0.936 &  0.683 &  -0.965 &  0.592 \\
   32) & 1.737 & 2.035 & -0.40 & -0.406 &  0.341 &   0.693 &  0.328 \\
   33) & 1.737 & 2.035 & -0.10 & -0.878 &  0.195 &   0.854 &  0.194 \\
   34) & 1.737 & 2.035 &  0.20 & -0.731 &  0.153 &   0.419 &  0.149 \\
   35) & 1.737 & 2.035 &  0.50 & -0.550 &  0.139 &   0.149 &  0.141 \\
   36) & 1.737 & 2.035 &  0.80 &  0.168 &  0.201 &   0.097 &  0.228 \\
   37) & 1.838 & 2.081 & -0.70 & -0.011 &  0.644 &  -0.303 &  0.579 \\
   38) & 1.838 & 2.081 & -0.40 & -1.000 &  0.507 &   1.215 &  0.386 \\
   39) & 1.838 & 2.081 & -0.10 & -0.891 &  0.240 &   0.319 &  0.232 \\
   40) & 1.838 & 2.081 &  0.20 & -0.912 &  0.179 &   0.524 &  0.178 \\
   41) & 1.838 & 2.081 &  0.50 & -0.429 &  0.164 &   0.048 &  0.148 \\
   42) & 1.838 & 2.081 &  0.80 & -0.745 &  0.253 &  -0.082 &  0.178 \\
   43) & 1.939 & 2.126 & -0.70 &  0.782 &  0.529 &   1.157 &  0.493 \\
   44) & 1.939 & 2.126 & -0.40 & -1.255 &  0.403 &   1.089 &  0.417 \\
   45) & 1.939 & 2.126 & -0.10 & -0.458 &  0.264 &   0.397 &  0.249 \\
   46) & 1.939 & 2.126 &  0.20 & -0.388 &  0.176 &   0.111 &  0.213 \\
   47) & 1.939 & 2.126 &  0.50 & -0.601 &  0.166 &   0.159 &  0.155 \\
   48) & 1.939 & 2.126 &  0.80 & -0.227 &  0.182 &  -0.193 &  0.164 \\
   49) & 2.039 & 2.170 & -0.70 & -1.268 &  0.480 &   1.279 &  0.444 \\
   50) & 2.039 & 2.170 & -0.40 & -1.606 &  0.430 &   0.700 &  0.434 \\
   51) & 2.039 & 2.170 & -0.10 & -0.629 &  0.294 &   0.958 &  0.284 \\
   52) & 2.039 & 2.170 &  0.20 & -1.315 &  0.196 &   0.235 &  0.208 \\
   53) & 2.039 & 2.170 &  0.50 &  0.004 &  0.188 &  -0.818 &  0.182 \\
   54) & 2.039 & 2.170 &  0.80 & -0.185 &  0.216 &  -0.134 &  0.170 \\
   55) & 2.139 & 2.212 & -0.70 & -0.019 &  0.457 &   0.047 &  0.429 \\
   56) & 2.139 & 2.212 & -0.40 & -0.535 &  0.438 &   0.098 &  0.471 \\
   57) & 2.139 & 2.212 & -0.10 & -0.405 &  0.332 &   0.800 &  0.314 \\
   58) & 2.139 & 2.212 &  0.20 & -0.507 &  0.214 &   0.462 &  0.219 \\
   59) & 2.139 & 2.212 &  0.50 & -0.029 &  0.183 &  -0.100 &  0.188 \\
   60) & 2.139 & 2.212 &  0.80 &  0.002 &  0.209 &  -0.041 &  0.163 \\
   61) & 2.240 & 2.255 & -0.70 &  1.186 &  0.761 &   0.828 &  0.649 \\
   62) & 2.240 & 2.255 & -0.40 &  0.134 &  0.515 &   1.377 &  0.600 \\
   63) & 2.240 & 2.255 & -0.10 & -0.920 &  0.404 &   0.783 &  0.440 \\
   64) & 2.240 & 2.255 &  0.20 & -0.424 &  0.300 &   0.294 &  0.310 \\
   65) & 2.240 & 2.255 &  0.50 & -0.312 &  0.203 &  -0.526 &  0.229 \\
   66) & 2.240 & 2.255 &  0.80 & -0.521 &  0.267 &  -0.170 &  0.245 \\
   67) & 2.341 & 2.296 & -0.70 & -0.074 &  0.676 &   1.019 &  0.603 \\
   68) & 2.341 & 2.296 & -0.40 &  0.010 &  0.686 &  -0.344 &  0.869 \\
   69) & 2.341 & 2.296 & -0.10 &  0.970 &  0.550 &  -0.064 &  0.565 \\
   70) & 2.341 & 2.296 &  0.20 & -0.435 &  0.398 &  -0.227 &  0.353 \\
   71) & 2.341 & 2.296 &  0.50 & -0.232 &  0.240 &  -0.442 &  0.257 \\
   72) & 2.341 & 2.296 &  0.80 & -0.042 &  0.227 &  -0.418 &  0.366 \\
   73) & 2.443 & 2.338 & -0.70 & -1.507 &  0.710 &   0.922 &  0.979 \\
   74) & 2.443 & 2.338 & -0.40 &  0.253 &  0.872 &   0.732 &  0.951 \\
   75) & 2.443 & 2.338 & -0.10 &  0.956 &  1.071 &  -1.261 &  0.881 \\
   76) & 2.443 & 2.338 &  0.20 &  0.137 &  0.581 &  -0.738 &  0.562 \\
   77) & 2.443 & 2.338 &  0.50 & -0.530 &  0.337 &  -0.308 &  0.383 \\
   78) & 2.443 & 2.338 &  0.80 & -0.015 &  0.327 &  -0.348 &  0.341 \\
   79) & 2.543 & 2.377 & -0.70 & -1.474 &  1.238 &   3.243 &  1.021 \\
   80) & 2.543 & 2.377 & -0.40 &  0.080 &  2.297 &   2.293 &  0.819 \\
   81) & 2.543 & 2.377 & -0.10 & -3.047 &  1.204 &   1.311 &  1.325 \\
   82) & 2.543 & 2.377 &  0.20 &  0.014 &  0.835 &  -1.254 &  0.847 \\
   83) & 2.543 & 2.377 &  0.50 &  0.459 &  0.363 &  -0.869 &  0.394 \\
   84) & 2.543 & 2.377 &  0.80 & -0.490 &  0.315 &   0.727 &  0.446 \\
   85) & 2.642 & 2.416 & -0.70 &  0.446 &  1.850 &   0.072 &  1.251 \\
   86) & 2.642 & 2.416 & -0.40 &  0.933 &  1.030 &   2.485 &  0.890 \\
   87) & 2.642 & 2.416 & -0.10 & -1.616 &  1.587 &  -2.157 &  1.191 \\
   88) & 2.642 & 2.416 &  0.20 & -1.441 &  0.948 &  -0.382 &  0.883 \\
   89) & 2.642 & 2.416 &  0.50 & -0.975 &  0.507 &  -0.207 &  0.476 \\
   90) & 2.642 & 2.416 &  0.80 &  0.587 &  0.413 &   0.181 &  0.797 \\
   91) & 2.741 & 2.454 & -0.70 & -0.986 &  1.066 &   0.999 &  1.074 \\
   92) & 2.741 & 2.454 & -0.40 & -1.042 &  1.022 &   1.256 &  1.224 \\
   93) & 2.741 & 2.454 & -0.10 &  1.461 &  1.218 &   0.891 &  1.477 \\
   94) & 2.741 & 2.454 &  0.20 & -0.333 &  1.151 &  -0.899 &  1.118 \\
   95) & 2.741 & 2.454 &  0.50 &  0.392 &  0.692 &  -1.106 &  0.504 \\
   96) & 2.741 & 2.454 &  0.80 & -0.743 &  0.505 &  -0.147 &  0.543 \\